\documentclass[12pt]{iopart}
\usepackage{graphicx}
\usepackage{comment}
\usepackage{iopams}
\usepackage{revsymb}
\usepackage{hyperref}
\usepackage{mathrsfs}
\newcommand{\fig}[2]{\includegraphics[width=#1]{#2}}

\begin{document}

\title{Exciton-assisted optomechanics with suspended carbon nanotubes}

\author{I~Wilson-Rae$^{1,2*}$, C.~Galland$^3$, W.~Zwerger$^2$,
  A.~Imamo\u glu$^3$} 
\address{$^1$Institute for Theoretical Physics, Universit\"at
  Erlangen-N\"urnberg, D-91058 Erlangen, Germany} 
\address{$^2$Technische Universit\"at M\"unchen, D-85748 Garching,
  Germany}  
\address{$^3$Institute of Quantum Electronics, ETH Z\"urich, Wolfgang-Pauli-Strasse 16, CH-8093 Z\"urich, Switzerland} 
%% Include email to one of the authors
\ead{${}^{*}$Ignacio.Wilson-Rae@ph.tum.de}

% \author{I.~Wilson-Rae} \email[]{ignacio.wilson-rae@ph.tum.de}
% \affiliation{Technische Universit\"{a}t M\"{u}nchen, 85748
%   Garching, Germany.}  
% \author{C.~Galland}
% \affiliation{Institute of Quantum Electronics, ETH Z\"urich,
%   Wolfgang-Pauli-Strasse 16, CH-8093 Z\"urich, Switzerland.} 
% \author{W.~Zwerger}
% \affiliation{Technische Universit\"{a}t M\"{u}nchen, 85748
%   Garching, Germany.}
% \author{A.~Imamo\u glu}
% \affiliation{Institute of Quantum Electronics, ETH Z\"urich,
%   Wolfgang-Pauli-Strasse 16, CH-8093 Z\"urich, Switzerland.} 
% \keywords{\ldots}

\date{\today}

\begin{abstract}
  We propose a framework for inducing strong optomechanical effects in
  a suspended carbon nanotube based on deformation potential
  exciton-phonon coupling. The excitons are confined using an
  inhomogeneous axial electric field which generates optically active
  quantum dots with a level spacing in the milli-electronvolt range
  and a characteristic size in the $10\,$nm range. A transverse field
  induces a tunable parametric coupling between the quantum dot and
  the flexural modes of the nanotube mediated by electron-phonon
  interactions. We derive the corresponding excitonic deformation
  potentials and show that this interaction enables efficient optical
  ground-state cooling of the fundamental mode and could allow us to
  realise the strong and ultra-strong coupling regimes of the
  Jaynes-Cummings and Rabi models.
\end{abstract}
%\pacs{85.85.+j, 42.50.Wk, 78.67.Ch, 63.22.Gh}
\maketitle

%keywords nanotubes,  optomechanics, exciton-confinement, cooling,
%quantum-dots, excitonic deformation-potential

%Original perpendicular field and (low) CNT rigidity 0.66\AA\times 1TPa
%Uncorrected deflection: 0.27nm
%Relative rigidity shift due to electrostatic softening: 0.2
%Corrected deflection neglecting nonlinearity: 0.335 nm

%(Original perpendicular field)/4 and revised CNT rigidity 340 N/m
%Uncorrected deflection: 0.0131nm
%Relative rigidity shift due to electrostatic softening: 0.00243
%Corrected deflection including nonlinearity: 0.0131nm
%Total relative frequency shift (elect. softening + nonlinearity): 0.00133
%Potential applied: -0.75 V / +1.75 V
%Field at z_0(=21.8 nm):  E_\perp = 0.045 V/nm
%Effective E_\perp : 0.0368 V/nm
%Max field at tip is approximately 0.2 V/nm (an upper and lower bound
%can be rigorously inferred from the relation to the original
%configuration that exactly determine the first significant figure)
%Confinement ZPM: 9 nm
%level spacing: 2.36 meV
%epsilonper = 1.6; epsilonpar = 7;
%mcm = 0.2 ElectronMass
%alphaa = 4 Pi VacuumPermittivity Angstrom^2;
%alphaper = 13 alphaa; alphapar = 75;
%s = 7.6 10^(-7) Kilogram/Meter^2;

\section{Introduction}

The realisation of the quantum regime of a macroscopic
mechanical degree of freedom \cite{Armour02,Blencowe04} has
emerged as a natural goal from considering the fundamental
limits for the measurement of small forces and displacements
\cite{Schwab05,Kippenberg08,Marquardt09,Favero09,Aspelmeyer12}. This
achievement could provide a versatile alternative for the
exploration of the quantum-to-classical transition and the
development of quantum technologies. Recent progress in the
implementation of optomechanical
\cite{Kippenberg08,Marquardt09,Favero09,Aspelmeyer12} and
electromechanical \cite{Schwab05} microsystems and nanosystems,
has already enabled access to this regime
\cite{O'Connell10,Rocheleau10,Teufel11a,Chan11,Verhagen12}. Some
critical requirements in this respect are: (i) quantum limited
measurement of the transduced output, (ii) sufficiently low
effective masses, and (iii) sufficiently low mechanical
dissipation. Optomechanical schemes are ideally suited to meet
(i) given the availability of shot-noise limited
photodetection. In turn suspended single-walled carbon nanotubes
(CNTs) \cite{Sazonova04} are emerging as unique candidates to
meet criteria (ii) and (iii) \cite{Tian04,Zippilli09}. Indeed
recent transport experiments in these systems have demonstrated
strong coupling of charge to vibrational resonances
\cite{Leturcq09,Steele09,Lassagne09} and mechanical
quality-factors ($Q$) exceeding $10^5$ for resonant frequencies
approaching $1\,$GHz \cite{Huttel09}. Thus, implementing
optomechanical systems based on suspended CNTs would represent a
significant milestone.

\begin{figure}[t]
  \fig{\linewidth}{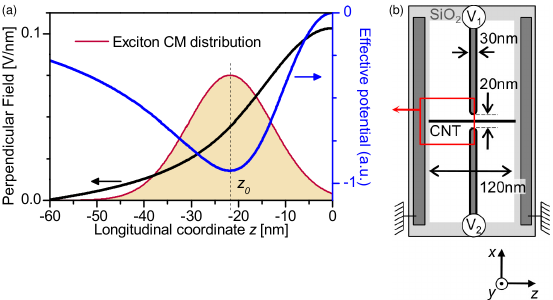}
  \caption{(a) Perpendicular field $E_\perp(z)$, effective
    confining potential $V_\mathrm{eff}^{(00)}(z)$, and
    resulting exciton CM probability distribution (a.u.) in the
    harmonic approximation for $V_1= -0.75\,$V and $V_2=
    1.75\,$V ---the tailored confinement generates a double well
    with minima at $|z|=z_0$
    [$E_\parallel(z_0)=9.6\mathrm{V}\mu^{-1}$ and inter-well
    tunnelling is negligible]. (b) Schematic view of the device
    (both the tip electrodes and the CNT are suspended). Here we
    consider the restrictions imposed on the fields by the need
    to avoid sizable static deflections
    (cf.~Sec.~\ref{subsec:implementation}), field emission (the
    maximum field at the tip electrodes is approximately
    $200\mathrm{V}\mu^{-1}$) and dielectric breakdown, and
    assume for simplicity cylindrical symmetry around the tips'
    axis.\label{fig:tips}}
\end{figure}

The standard paradigm in optomechanics is based on an optical
cavity whose frequency is modulated by the motion of one of its
mirrors or of a dielectric object inside it via radiation
pressure effects
\cite{Kippenberg08,Marquardt09,Favero09}. Though recently this
approach has been successfully carried over to nanoresonators
with subwavelength dimensions
\cite{Favero09b,Anetsberger09,Favero08}, it becomes inefficient
for high-frequency resonators ($\gtrsim1\,$GHz) with low
polarisabilities like sub-micron CNTs. Here we propose a
promising alternative based on a way of inducing coherent
optomechanical transduction that exploits the unique properties
of excitons in CNTs
\cite{Htoon04,Hogele08,Torrens08,Srivastava08}. The role of the
optical cavity is played by an excitonic resonance of the CNT
that couples parametrically to the flexural motion via
deformation potential electron-phonon interactions
\cite{Suzuura02,Mariani09}.  Homodyne detection of the in-phase
component of the output field of the resonantly driven two-level
emitter afforded by the excitonic resonance allows then to
perform a continuous measurement of the mechanical
displacement. This procedure, which could be implemented using
the differential transmission technique \cite{Hoegele04}, is
akin to ion-trap measurements \cite{Gerritsma10} and analogous
to cavity-assisted schemes \cite{Clerk10}. However a two-level
emitter is a highly nonlinear system, in stark contrast to an
optical cavity mode, and this feature provides an additional
motivation for our proposal given the potential it offers for
the demonstration of non-classical motional states of the
CNT. In fact, a major advantage of this particular mechanical
resonator-exciton system with respect to prior scenarios
\cite{Wilson-Rae04} is the possibility of realising a mechanical
analogue of the strong-coupling regime of cavity-QED
\cite{Thompson92,Gleyzes07} with a vacuum Rabi splitting in the
$100\,$MHz range.

As shown in Fig.~\ref{fig:tips}, we envisage a suspended neutral
semiconducting CNT where the centre of mass (CM) of the exciton
is localised via the spatial modulation of the Stark-shift
induced by a static inhomogeneous electric field.  We analyse a
tip-electrode configuration that effectively engineers a pair of
tunable optically active nanotube quantum dots (NTQDs) with
excitonic level spacing in the meV range implying a confinement
length below $10\,$nm.  The quantum confinement is induced by
the inhomogeneity in the field component along the CNT axis
$E_\parallel$. In turn the normal component $E_\perp$ breaks the
axial symmetry inducing a tunable parametric coupling between
the exciton and the in-plane flexural motion of the CNT. This
allows for optical ground-state cooling
\cite{Courty01,Vitali02,Wilson-Rae04,Chan11} of the fundamental
mode at an ambient temperature in the Kelvin range.

Another aspect of the CNT systems proposed here is their
relevance in the broader context of quantum photonics
applications.  The optical manipulation of single carriers
trapped at well-defined and repeatedly addressable locations has
proved invaluable to probe the foundations of quantum mechanics
and constitutes a key resource for enabling quantum
technologies. % \cite{Leibfried03}
In this respect solid state zero-dimensional optical emitters
offer a promising alternative to realisations based on trapped
atoms. Examples of such systems are semiconductor quantum dots
\cite{Petroff01}, rare-earth doped crystals
\cite{Ohlsson03,deRiedmatten08}, NV colour centres in diamond
\cite{Jelezko03,Gaebel06} and, more recently, the system
considered here, i.e.~excitons in semiconducting CNTs
\cite{Hogele08}.  For the latter photon antibunching has already
been observed and there is indirect evidence that exciton
localisation arises spontaneously \cite{Hogele08}, probably as a
result of charged defects in the surrounding matrix or disorder
in the CNT. Naturally, in view of the aforementioned quantum
applications it would be desirable to artificially tailor
exciton confinement in a controlled fashion in an ultra-clean
suspended CNT.  However the band-gap engineering approach
deployed for this purpose in more standard semiconductor
materials is hardly viable for CNTs. Suspension is motivated by
the need to (i) avoid substrate-induced fluorescence quenching
and (ii) control the enhanced phonon-induced optical dephasing
that results from the one-dimensional nature of this system.

We will first present in Sec.~\ref{sec:Xmodel} a suitable model
for the CNT excitons that will be used to analyse both their
quantum confinement generating the NTQD and their interaction
with CNT low-frequency phonons, as modified by the applied
electrostatic fields. Subsequently, in
Sec.~\ref{sec:Xoptomechanics}, we will consider the axial phonon
confinement induced by the finite CNT suspended length and
derive an effective model for the fundamental in-plane
flexural-phonon mode coupled to the NTQD. Finally, also in
Sec.~\ref{sec:Xoptomechanics}, this effective model will be used to
describe optomechanical effects induced when the NTQD is driven
by a laser --- e.g.~the possibility of efficient laser cooling
of the fundamental mode close to its motional quantum ground
state.

\section{Exciton effective model}\label{sec:Xmodel}

The electronic structure of a CNT can be understood in terms of
graphene rolled into a seamless cylinder. In graphene, the conduction and
valence bands cross at two inequivalent ``Dirac points'' in the
reciprocal lattice \cite{DiVincenzo84} ($K$ and $K'$). The
corresponding effective ($k\cdot p$) Dirac Hamiltonian reads
\begin{equation}\label{dirac}
  H_D=\hbar v_F \left(k_{x'} \hat{\sigma}_{x'} + k_{y'}
    \hat{\tau}_3\hat{\sigma}_{y'} \right)\,, 
\end{equation} 
where $\hat{\sigma}_{i}$ [$\hat{\tau}_{i}$] are Pauli matrices in
sublattice ($A$-$B$) space [valley ($K$-$K'$) space] and $v_F\approx
10^6\,$m$\,$s$^{-1}$ %1.195 Araujo07
is the Fermi velocity. Here we follow the representation used by Ando
\cite{Ando06} % in
% Ref.~\onlinecite{Ando06}
in which the Dirac spinor is expressed in terms of Bloch
amplitudes as
$(\psi_A^{K},\psi_B^{K},\psi_A^{K'},-\psi_B^{K'})$.  A rotation
by the CNT's chiral angle $\theta$ is performed to align
$\hat{y}'$ with the nanotube axis
($\hat{y}'\to\hat{z}$) ---henceforth, we adopt cylindrical
coordinates. %  and drop the primes of the Pauli matrices'
% indices.
Thus, the unit vectors in sublattice and valley space correspond
to graphene 2D Bloch functions with wavevector $K$ [$K'$]:
$(1,0,0,0)$ [$(0,0,1,0)$] is constructed from $p_z$ orbitals
centred at the $A$ sites with phase $e^{-i\theta/2}$
[$e^{i\theta/2}$] at the central site, while $(0,1,0,0)$
[$(0,0,0,1)$] has the $p_z$ orbitals centred at the $B$ sites
with phase $e^{i\theta/2}$ [$-e^{-i\theta/2}$] at the central
site \cite{DiVincenzo84}. After folding, the resulting periodic
boundary conditions imply that a wide-gap semiconducting
nanotube is obtained when the nanotube indices satisfy
mod$(2n'+m',3)\neq0$. The associated subband electronic 1D Bloch
functions % at the Dirac points
for $k_z=0$ are given by
\begin{eqnarray}\label{states}
  \!\!\!\!\!\!\!\!\!\!\!\!\!\!\!\!\!\!\!\!\!\!\!\!\!\!\!\!\!\!\!\!\!\!
  |K_{n,\pm}\rangle\doteq  
  \frac{e^{i(n-\nu/3)\varphi}}{\sqrt{4\pi}} 
%\begin{pmatrix}
\left(\begin{array}{c}
  \mathrm{sgn} [3n-\nu]\\ \pm1\\0\\0
%\end{pmatrix}
\end{array}\right)
\,, \qquad |K'_{n,\pm}\rangle  \doteq \frac{e^{i(n+\nu/3)\varphi}}{\sqrt{4\pi}}
% \begin{pmatrix}
\left(\begin{array}{c}
   0\\0\\ \mathrm{sgn} [3n+\nu] \\ \pm1
% \end{pmatrix}
\end{array}\right)
\,,
\end{eqnarray}
with single-particle eigenenergies 
\begin{equation}\label{energies}
  \epsilon_{n,\pm}=\pm\frac{\hbar
    v_F}{R}\left|n-\frac{\nu}{3}\right|\,,\qquad\epsilon'_{n,\pm}=\pm\frac{\hbar
    v_F}{R}\left|n+\frac{\nu}{3}\right|\,. 
\end{equation}
Here $+$ and $-$ label, respectively, the conduction and valence
bands, the index $n$ corresponds to the quantisation of the
single-particle azimuthal momenta ($n=0,\pm1,\pm2,\ldots$),
$\nu=2$mod$(2n'+m',3)-3=\pm1$ denotes the type of
semiconducting CNT \cite{Jorio05}, $R$ is the CNT radius, and
$\varphi$ is the azimuthal angular coordinate
(cf.~Fig.~\ref{fig:tips}).

The $K-K'$ degeneracy, implied by Eqs.~(\ref{states}) and
(\ref{energies}), and the electron spin yield for the
lowest-energy exciton manifold a sixteen-fold degeneracy in the
lowest order $k\cdot p$ approximation provided by
Eq.~(\ref{dirac}). This degeneracy is partially split by
inter-valley mixing and exchange effects \cite{Ando06}.  In the
absence of a magnetic field there is a single bright
level\footnote{Note however that manipulating the indirect
  excitons may also be viable \cite{Torrens08}.}: the singlet
bonding direct exciton $|KK^*\rangle+|K'{K'}^{*}\rangle$ which
typically has the highest energy \cite{Chang04} (here the
conjugated wavefunctions correspond to the hole).  Threading
through the cross section a small Aharonov-Bohm flux
$\phi_\mathrm{AB}$ renders the antibonding state
$|KK^*\rangle-|K'{K'}^{*}\rangle$ weakly allowed \cite{Ando06}
so that its spontaneous emission rate % $\Gamma_-$
can be tuned by varying the magnetic field. Note that the zero
field splitting between the states
$|KK^*\rangle\pm|K'{K'}^{*}\rangle$ % lies
has been observed to be in the $1\,$meV range
\cite{Srivastava08}.

Henceforth, we will mainly focus on the $E_{11}$ excitons
$|KK^*\rangle\pm|K'{K'}^{*}\rangle$ and consider their deformation
potential (DP) coupling to the low-frequency acoustic phonons of the
CNT in the presence of weak static electric fields. There are four
phonon branches whose frequency vanishes with the
wavevector, % with no infrared
% cutoff, 
namely: compressional (stretching), torsional and two flexural
(bending) branches. However relevant exciton-phonon interactions
will only affect the compressional branch and the flexural
branch polarised along $\hat{x}$. Indeed, the excitonic
wavefunctions of interest are non-degenerate and, in the absence
of a magnetic field (i.e.~for $\phi_\mathrm{AB}=0$), the field
configuration in Fig.~(\ref{fig:tips}) has reflection symmetry
with respect to the $xz$ plane. Thus, the torsional branch and
the flexural branch polarised along $\hat{y}$, which are
antisymmetric with respect to this reflection operation, are
decoupled. % \footnote{For finite flux the symmetries also imply
  % that the correction, due to the magnetic field, to the DP
  % affecting the other branches is at most quadratic in
  % $\phi_\mathrm{AB}$.}.

To obtain a tractable model for the excitonic wavefunction suitable
for analysing the CM confinement of the exciton and its coupling to
acoustic phonons, we adopt a phenomenological approach based on: (i)
the aforementioned $k\cdot p$ graphene zone-folded scheme following
Ando \cite{Ando06,Uryu06} but neglecting inter-subband transitions
(cf.~Refs.~\cite{Ando04,Capaz06}), and (ii) an envelope function
approximation within each subband with the parametrisation developed
in Refs.~\cite{Jorio05,Capaz06} (based on a comparison of tight
binding and \emph{ab initio} approaches) but the Bloch function at
$K$, $K'$ as determined by (i) and assuming electron-hole symmetry.
In line with this last assumption we take for the effective mass
associated to a given subband the average of the electron and hole
effective masses obtained in Ref.~\cite{Jorio05}. This framework
neglects the impact of curvature effects on the Bloch functions
% \footnote{Here the Bloch functions mainly play a role in determining
%   the selection rules.}
but allows to incorporate realistic values for the effective masses of
electrons and holes \cite{Jorio05} and for the exciton Bohr radius
\cite{Capaz06} ---here the Bloch functions mainly play a role in
determining the selection rules. In the absence of external fields,
this leads to the following singlet direct exciton wavefunctions
\begin{eqnarray}\label{wavefunction}
  |\psi_{nm\pm}\rangle = &\,
  \frac{1}{2}\left(|K_{n,+}K_{m,-}^{*}\rangle\otimes|F_{nm}\rangle
    \pm
    |K'_{-m,+}{K'}_{-n,-}^{*}\rangle\right.\nonumber\\
  &\left.\otimes|F'_{nm}\rangle\right)\otimes
  \left(|\uparrow\downarrow\rangle+
    |\downarrow\uparrow\rangle\right)\,,
\end{eqnarray}
where the envelope functions $|F_{nm}\rangle$, $|F'_{nm}\rangle$
satisfy $F_{nm}(z_e,z_h)=F'_{nm}(z_h,z_e)$. Here we use that the
single-particle eigenenergies $\epsilon_{n,+}$
($\epsilon_{n,-}$) and $\epsilon'_{-n,+}$ ($\epsilon'_{-n,-}$)
are degenerate and inter-valley mixing preserves the total
orbital momentum and angular momentum along $\hat{z}$. One
should note that in general Eq.~(\ref{wavefunction}) implies
entanglement between the valley and orbital degrees of freedom
but for axially-symmetric excitons such that $n=m$, the envelope
functions differ at most by a $\pi$-phase so that they factor
out and the axial degrees of freedom become disentangled
($n=m=0$ for $E_{11}$ excitons).

\subsection{Exciton confinement}\label{subsec:Xconfinement}

We turn now to the analysis of the quantum confinement of the
exciton's CM motion induced by an applied inhomogeneous
electrostatic field. The symmetric tip electrode configuration
sketched in Fig.~\ref{fig:tips} with voltages $V_1$ and $V_2$
allows independent tuning of $E_\perp$ and $E_\parallel$ as the
reflection symmetries imply that their magnitudes are
determined, respectively, by $(V_2-V_1)/2$ and $(V_2+V_1)/2$. In
order to generate a Stark-shift induced excitonic quantum dot,
we consider parameters such that the length scale over which
$E_\perp$ and $E_\parallel$ vary appreciably is much larger than
the CNT radius $R$ (i.e.~they can be regarded as constant across
the CNT's cross section) and larger than the excitonic Bohr
radius. It follows that for sufficiently weak magnitudes (see
below): (i) the effect of $E_\parallel$ is dominated by
intrasubband virtual transitions whose effect on the CM motion
can be treated adiabatically, while (ii) $E_\perp$ leads to a
perturbation $\propto e^{\pm i\varphi}$ that only induces
intersubband virtual transitions.

More precisely, the effect of the external fields on the exciton
can be described by the following interaction Hamiltonian
\begin{equation}\label{V}
\hat{V} = \hat{V}_\parallel + \hat{V}_\perp
\end{equation}
with 
\begin{equation}\label{Vparallel}
  \hat{V}_\parallel = e \left[U(0,\hat{z}_h)-U(0,\hat{z}_e)\right]
\end{equation}
and
\begin{equation}\label{Vperp}
  \hat{V}_\perp = \frac{e}{\epsilon_\perp}
  \left[\hat{x}_e\, E_\perp(\hat{z}_e) -\hat{x}_h\, E_\perp(\hat{z}_h)\right]\,. 
\end{equation}
Here $U(x,z)$ is the external electrostatic potential (in the
plane $y=0$), % along the
% CNT axis,
so that $E_\parallel(z)=-\frac{\partial U}{\partial z}(0,z)$ and
$E_\perp(z)=-\frac{\partial U}{\partial x}(0,z)$, and
$\epsilon_\perp\approx1.6$ denotes the intrinsic relative
permittivity normal to this axis \cite{Fagan07a}. Given that the
depolarisation effect only affects the perpendicular field and
the fact that the typical level spacing between the relevant
excitonic states arising from a given pair of subbands $n,\,m$,
is smaller than the intersubband energy spacing, % and the
% excitonic Bohr radius is larger than the CNT radius (the ratio
% of the energy shifts induced by $\hat{V}_\parallel$ and
% $\hat{V}_\perp$ is $\sim 30$)
we can assume that the Stark-shift induced CM confinement will
be dominated by $\hat{V}_\parallel$ and neglect $\hat{V}_\perp$
in this subsection \footnote{More precisely, for
  $E_\parallel(z_0)\sim E_\perp(0)$ the relative order of the
  transverse field's contribution to the effective confining
  potential would be given by
  $\xi\mathcal{E}_*\frac{R}{\sigma_{eh}}\sim10^{-3}$ for
  relevant nanotube parameters (cf.~Eq.~(\ref{adiabatic}) and
  Secs.~\ref{subsec:X-phonon} and
  \ref{subsec:laser-cool}).}. Then, for each pair of subbands
$n,\,m$, we perform the substitution
\begin{equation}\label{sust}
z_{e/h}=z_\mathrm{CM}\mp\frac{\mu}{m^{(n/m)}}\,z'
\end{equation}
with inverse
\begin{equation}\label{sust_inv}
  z_\mathrm{CM}=\frac{m^{(n)}z_e+ m^{(m)}z_h}{m^{(n)}+ m^{(m)}}\,,
  \qquad  z'=z_h-z_e\,,
\end{equation}
where $m^{(n)}$ ($m^{(m)}$) is the effective mass of subband $n$ ($m$)
and $\mu=m^{(n)}m^{(m)}/(m^{(n)}+ m^{(m)})$ is the excitonic reduced
mass. Here the indices $n,\,m$ refer to the $K$ point and the
corresponding envelope function is enough to determine the
wavefunction [cf.~Eq.~(\ref{wavefunction})] ---note that an analogous
substitution can be performed for the $K'$ point and
$m'{}^{(-n)}=m^{(n)}$.  The aforementioned smoothness of the fields
implies $\langle \hat{z}_\mathrm{CM}^2\rangle\gg\langle
\hat{z}'{}^{2}\rangle$ so that substituting Eq.~(\ref{sust}) into
Eq.~(\ref{Vparallel}) and Taylor expanding we obtain a linear
potential
\begin{equation}\label{Vparapprox}
  \hat{V}_\parallel = e \hat{z}'\frac{\partial U}{\partial
    z}(0,\hat{z}_\mathrm{CM}) + \mathcal{O} \left[\hat{z}'{}^{2}\right]\,. 
\end{equation}
 
We now consider $E_\parallel(z)$ much smaller than the critical field
to ionise the exciton so that the relevant intrasubband matrix
elements of $\hat{V}_\parallel$ are much smaller than the binding
energy. The latter warrants for the ground state manifold of the
excitonic hydrogenic series \cite{Barros06} associated to $n,\,m$, a
description in terms of an effective Hamiltonian by adiabatic
elimination of the corresponding excited manifolds. Naturally, the
corresponding states have well-defined parity. Thus, this effective
Hamiltonian for the CM motion has a potential part whose leading
contribution is second order in $\hat{V}_\parallel$. From
Eq.~(\ref{Vparapprox}) we obtain for this effective confining
potential
\begin{equation}\label{Veff}
  V_\mathrm{eff}^{(nm)}(z_\mathrm{CM})=-\alpha_X^{(nm)}E_\parallel^2(z_\mathrm{CM})\,,  
\end{equation}
where
\begin{equation}\label{alphaX}
  \alpha_X^{(nm)}=e^2 \sum_{l\neq0} \frac{\left|\langle
      f_{eh,0}^{(nm)}|\hat{z}'|f_{eh,l}^{(nm)}\rangle\right|^2}{E^{(nm)}_l-E^{(nm)}_0} 
\end{equation}
is the corresponding excitonic polarisability. Here we have introduced
the wavefunctions $\{|f_{eh,l}^{(nm)}\rangle\}$ on the relative
coordinate $z'$ for the hydrogenic series $l=0,1,\ldots$ arising from
$n,\,m$. In a classical picture, the field polarises the exciton that
thus experiences a force proportional to the gradient of the squared
field.  If one considers the characteristic level spacings
\begin{equation}\label{spacings}
E^{(nm)}_l-E^{(nm)}_0\sim \frac{e^2}{8\pi\epsilon_0\epsilon_{n-m}(0)\sigma_{eh}^{(nm)}}
\end{equation} 
and the order of the relevant matrix elements $\sim
\sigma_{eh}^{(nm)}$ (where $\sigma_{eh}^{(nm)}$ is the exciton Bohr radius and
$\epsilon_{n-m}(q)$ an appropriate dielectric function
\cite{Ando06}), it follows from Eq.~(\ref{alphaX}) that the
excitonic polarisability is of order
\begin{equation}\label{Xpolarizability} 
  \alpha_X^{(nm)}\!\sim\!8\pi\epsilon_0 \epsilon_{n-m}(0){\sigma_{eh}^{(nm)}}^3\,.
\end{equation}
Here $\epsilon_{0}(0)=\epsilon_\parallel\approx7$ corresponds to the
intrinsic relative permittivity along the CNT axis \cite{Fagan07a}.

We focus now on $E_{11}$ excitons so that $n=m=0$.  To determine
the electrostatic potential generated by the aforementioned
symmetric tip electrode configuration, we have performed FEM
calculations with COMSOL multiphysics for relevant parameters
(cf.~Sec.~\ref{subsec:laser-cool}). We assume for simplicity
cylindrical symmetry around the tips' axis so that the ground
electrode corresponds to a cylindrical shell with diameter
approximately equal to the CNT length $L$
(cf.~Fig.~\ref{fig:tips}). To model this configuration with
electrodes of length $L_T\gg L$ we consider Neumann boundary
conditions at $|x|=L_*$ with $L_*$ smaller but comparable to
$L_T$ (i.e.~$E_x=0$ for $|x|=L_*\sim1\,\mu$). The effective
confining potential $V_\mathrm{eff}$ is then estimated from
Eqs.~(\ref{Veff}) and (\ref{Xpolarizability}). As expected it
vanishes at the origin and is given by a double well with minima
at $|z|=z_0$, comparable to the distance between the tip
electrodes (i.e.~the tailored confinement generates a QD
molecule). For typical parameters, it is permissible to neglect
inter-well tunnelling and use in our estimates the harmonic
approximation for the ground state of each well. We consider as
a representative example a $(9,4)$ CNT,
$E_\parallel(z_0)=9.6\,\mathrm{V}\mu^{-1}$ and
$z_0=21\,\mathrm{nm}$ (cf.~Fig.~\ref{fig:tips}).  This results
in an excitonic CM zero point motion
$\sigma_\mathrm{CM}^{(00)}\sim9\,\mathrm{nm}$ (corresponding to
$\sigma_\mathrm{eh}^{(00)}\approx1.5\,\mathrm{nm}$
\cite{Capaz06} and $m_\mathrm{CM} = 0.2 m_e$ \cite{Jorio05},
with $m_e$ the electron mass in vacuum) and a level spacing
$\Delta E\sim2\,\mathrm{meV}$ ---where
$\frac{\partial^2E^2_\parallel}{\partial z^2}(z_0)$ is taken
from the FEM calculation. Alternatively, heuristic
considerations imply that % In
% general
% a suitable heuristic estimate of
the curvature at the well minima is % afforded by
of order $|V_\mathrm{eff}^{(nm)}(z_0)|/z_0^2$ which, together
with Eqs.~(\ref{Veff}) and (\ref{Xpolarizability}), yields
\begin{equation}\label{sigmaCM}
  \sigma_\mathrm{CM}\sim\frac{\sqrt{\hbar z_0/E_\parallel(z_0)}}{2\left[2\pi
      m_\mathrm{CM} \epsilon_0\epsilon_\parallel
      {\sigma_\mathrm{eh}}^3\right]^{1/4}}\, 
\end{equation}
and
\begin{equation}\label{DeltaE}
  \Delta E\sim \frac{\hbar E_\parallel(z_0)}{2
    z_0}\sqrt{\frac{2\pi\epsilon_0\epsilon_\parallel 
      {\sigma_\mathrm{eh}}^3}{m_\mathrm{CM}}}\,  
\end{equation}
where we have omitted the subband superscripts. Given the
subwavelength separation between the wells, for each doublet
only the symmetric state is bright so that the degeneracy is
irrelevant. Thus, the wavefunctions for relevant $E_\parallel$
are well approximated by the form given in
Eq.~(\ref{wavefunction}) with a suitable envelope function
\begin{equation}\label{envelope_F}
F_{nm}(z_e,z_h)=f_{eh}^{(nm)}(z_h-z_e)
f_\mathrm{CM}(\textstyle{\frac{m^{(n)}z_e+ m^{(m)}z_h}{m^{(n)}+ m^{(m)}}}) 
\end{equation}
[cf.~Eq.~(\ref{sust_inv})], and for the ground state of the
$E_{11}$ bonding (antibonding) 
manifold, henceforth denoted by $|\psi_{00+}\rangle$
($|\psi_{00-}\rangle$), we take both functions $f_\mathrm{CM}$
and $f_{eh}$ to be Gaussian \cite{Capaz06}.

Finally, more precise criteria to warrant the above adiabatic
treatment are given by
\begin{equation}\label{adiabatic}
  \rm{(i)}\quad \  |E_\parallel(z_0)|\ll
  e/8\pi\epsilon_0\epsilon_\parallel\sigma_\mathrm{eh}^2\equiv\mathcal{E}_* \qquad \
  \rm{and}\qquad \ \rm{(ii)}\quad \ z_0\gg\sigma_\mathrm{eh}\,; 
\end{equation}
which follow, respectively, from requiring that the relevant
matrix elements of $\hat{V}_\parallel$ be much smaller than the
corresponding unperturbed energy differences
[cf.~Eq.~(\ref{spacings})] and that the variation of the fields
be smooth enough \footnote{We note that $\mathcal{E}_*$ is
  comparable to the axial critical field for exciton
  dissociation and yields $\sim50\,$V$\mu^{-1}$ for relevant CNT
  parameters (cf.~Sec.~\ref{subsec:laser-cool}).}. This second
condition (ii) is necessary, together with (i), to ensure
\begin{equation}\label{consistency}
  \Delta E\ll E^{(nm)}_l-E^{(nm)}_0\quad\rm{or\ alternatively}\quad
  \sigma_\mathrm{CM}\gg\sigma_\mathrm{eh} \,, 
\end{equation}
which justifies the Taylor expansion leading to Eq.~(\ref{Vparapprox}),
and can be obtained using Eqs.~(\ref{Veff}), (\ref{spacings}),
(\ref{Xpolarizability}), (\ref{sigmaCM}) and (\ref{DeltaE}) together with the
assumptions $\sigma_{eh}\sim 4\pi\epsilon_0\epsilon_\parallel\hbar^2
/\mu e^2$ and $\mu\sim m_\mathrm{CM}$.

\subsection{Field-induced exciton-phonon coupling}\label{subsec:X-phonon}

In what follows we will first consider the influence of the
transverse field $E_\perp$ on the ground state of the $E_{11}$
axially-polarised exciton, which will amount to a slight
hybridisation with the cross-polarised excitonic states, and
then proceed to derive the corresponding excitonic deformation
potentials in the presence of the applied fields. 

To this end we treat the effect of $\hat{V}_\perp$ on
$|\psi_{00\pm}\rangle$ to lowest order in perturbation
theory. In principle the linear correction
$|\psi_{00\pm}^{(1)}\rangle$ involves contributions from all
four excitonic manifolds for which $|n|=1$, $m=0$ or $n=0$,
$|m|=1$, namely $E_{12}$, $E_{21}$, $E_{13}$ and $E_{31}$. It is
straightforward to determine the necessary single-particle
matrix elements of $\hat{x}$ within our model for the excitonic
wavefunctions, provided that the 2D ``lattice contributions''
(on the folded graphene sheet) are first reduced to momentum
matrix elements which can be read out from the $k\cdot p$
Hamiltonian (\ref{dirac}). To this effect one can first consider
the CNT as a 1D lattice so that there is only a ``lattice
contribution'' to the matrix elements.  Given that the operators
$\hat{x}$ and $\hat{p}_x$ are well-behaved on the corresponding
Bloch functions (which have finite range in the transverse
directions) we can apply the standard relation between their
matrix elements
\begin{equation}\label{xp:relation}
  \langle K_{m,\pm}| \hat{x}|K_{n,\pm}\rangle=\frac{i\hbar\langle K_{m,\pm}|
    \hat{p}_x|K_{n,\pm}\rangle}{m_e \left(\epsilon_{n,\pm}-\epsilon_{m,\pm}\right)}
\end{equation}
reducing the problem to the evaluation of matrix elements of
$\hat{p}_x$ \footnote{Note that the validity of
  Eq.~(\ref{xp:relation}) hinges on the fact that homogeneous
  boundary conditions are satisfied for $x,y\to\pm\infty$, and
  this relation between matrix elements breaks down for periodic
  boundary conditions as those satisfied along $z$.}  ---
analogous relations hold for states at $K'$. We now use: (i)
$\hat{p}_x=\cos\!\hat{\varphi}\,\hat{p}_r-\sin\!\hat{\varphi}\,\hat{p}_\varphi$,
(ii) that $\Pi$-orbitals have reflection symmetry with respect
to the graphene sheet so that they do not carry $\hat{p}_r$, and
(iii) that for Bloch functions near the Dirac points
$\hat{p}_\varphi\to m_e v_F\hat{\sigma}_{x'}-\frac{i\hbar}{R}
\frac{\hat{\partial}}{\partial\varphi}$. These, together with
Eqs.~(\ref{energies}) and (\ref{xp:relation}) imply
%\begin{widetext}
\begin{equation}\label{matrix-x}
  \langle K_{m,\pm}|\hat{x}|K_{n,\pm}\rangle= \frac{\mp i
    \langle
    K_{m,\pm}|\sin\!\hat{\varphi}\left(R\hat{\sigma}_{x'} -
    \frac{i\hbar}{m_ev_F}\frac{\hat{\partial}}{\partial\varphi}\right)
    |K_{n,\pm}\rangle}{|n-\nu/3| 
    -|m-\nu/3|}\,.
\end{equation}
%\end{widetext}
Subsequently, from Eqs.~(\ref{states}), (\ref{energies}) and
(\ref{matrix-x}), and using that $\nu=\pm1$ implies
\begin{eqnarray}\label{steps}
  &\mathrm{sgn} [3n-\nu]|_{n=\mp\nu} =\mp\nu\,, \quad
  \mathrm{sgn} [3n-\nu]|_{n=0}  =-\nu \,,\nonumber\\
  &\frac{i}{2\pi}\int_0^{2\pi}\!\rmd\varphi\, e^{i
    \nu\varphi}\sin\varphi = -\frac{\nu}{2}\,, \quad
  \int_0^{2\pi} \!\frac{\rmd\varphi}{2\pi}\,
  e^{i\frac{4\nu}{3}\varphi}\sin\varphi
  \,\frac{\partial}{\partial\varphi}e^{-i\frac{\nu}{3}\varphi}=
  \frac{1}{6}\,,
\end{eqnarray}
%\end{widetext}
and that % $|K_{n,\pm}\rangle$ are eigenstates of
% $\hat{\sigma}_{x'}$;
\begin{equation}\label{eigen}
  \hat{\sigma}_{x'}|K_{0,\pm}\rangle=\mp\nu|K_{0,\pm}\rangle\,,
  \qquad \hat{\sigma}_{x'}|K_{-\nu,\pm}\rangle=\mp\nu|K_{-\nu,\pm}\rangle\,;
\end{equation}
we obtain
% \begin{eqnarray}\label{matrix-x2}
% \langle K_{-\nu,\pm}|\hat{x}|K_{0,\pm}\rangle & =\frac{R}{2} \pm
% \frac{\hbar}{6 m_ev_F}\,, \nonumber\\
% \langle K_{+\nu,\pm}|\hat{x}|K_{0,\pm}\rangle & = 0 \,.
% \end{eqnarray}
\begin{equation}\label{matrix-x2}
\langle K_{-\nu,\pm}|\hat{x}|K_{0,\pm}\rangle  =\frac{R}{2} \pm
\frac{\hbar}{6 m_ev_F}\,, \qquad
\langle K_{+\nu,\pm}|\hat{x}|K_{0,\pm}\rangle  = 0 \,.
\end{equation}
% \end{widetext}
% \begin{eqnarray}\label{steps}
%   \mathrm{sgn} [3n-\nu]|_{n=\pm\nu} =\pm\nu\,, & \qquad
%   \mathrm{sgn} [3n-\nu]|_{n=0}  =-\nu \,,\nonumber\\
%   \frac{i}{2\pi}\int_0^{2\pi}\rmd\varphi e^{i
%     \nu\varphi}\sin\varphi = -\frac{\nu}{2}\,, & \qquad
%   \int_0^{2\pi} \frac{\rmd\varphi}{2\pi}
%   e^{i\frac{4\nu}{3}\varphi}\sin\varphi
%   \frac{\partial}{\partial\varphi}e^{-i\frac{\nu}{3}\varphi}=
%   \frac{1}{6}\,,
% \end{eqnarray}
% and that $|K_{n,\pm}\rangle$ are eigenstates of
% $\hat{\sigma}_{x'}$; we obtain
In turn, Eqs.~(\ref{states}) and (\ref{energies}) imply that the
substitution $\nu\to -\nu$ yields for the states
$|K'_{n,\pm}\rangle$ and $|K'_{m,\pm}\rangle$, expressions
analogous to Eqs.~(\ref{matrix-x}), (\ref{steps}), (\ref{eigen})
and (\ref{matrix-x2}) so that
% \begin{eqnarray}\label{matrix-x2prime}
% \langle K'_{+\nu,\pm}|\hat{x}|K'_{0,\pm}\rangle & =\frac{R}{2} \pm
% \frac{\hbar}{6 m_ev_F}\,, \nonumber\\
% \langle K'_{-\nu,\pm}|\hat{x}|K'_{0,\pm}\rangle & = 0 \,.
% \end{eqnarray}
\begin{equation}\label{matrix-x2prime}
\langle K'_{+\nu,\pm}|\hat{x}|K'_{0,\pm}\rangle  =\frac{R}{2} \pm
\frac{\hbar}{6 m_ev_F}\,, \qquad
\langle K'_{-\nu,\pm}|\hat{x}|K'_{0,\pm}\rangle  = 0 \,.
\end{equation}
Finally, from Eqs.~(\ref{wavefunction}), (\ref{Vperp}),
(\ref{matrix-x2}), and (\ref{matrix-x2prime}), and using that
$\hat{x}$ is invariant under time reversal; it is
straightforward to obtain
\begin{eqnarray}
 & \!\!\!\!\!\!\!\!\!\!\!\!\!\!\!\!\!\!\!\!\!\!\!
\langle\psi^{(0)}_{-\nu0\mp}|\hat{V}_\perp|\psi^{(0)}_{00\pm}\rangle 
  =-\langle\psi^{(0)}_{0-\nu\mp}|\hat{V}_\perp|\psi^{(0)}_{00\pm}\rangle=
  \frac{e R}{2\epsilon_\perp} \langle
  F_{-\nu0}|E_\perp(\hat{z}_e)|F_{00}\rangle\,, \nonumber \\
 & \!\!\!\!\!\!\!\!\!\!\!\!\!\!\!\!\!\!\!\!\!\!\!
\langle\psi^{(0)}_{-\nu0\pm}|\hat{V}_\perp|\psi^{(0)}_{00\pm}\rangle  =
  \langle\psi^{(0)}_{0-\nu\pm}|\hat{V}_\perp|\psi^{(0)}_{00\pm}\rangle=
  \frac{\hbar e}{6 \epsilon_\perp m_ev_F} \langle
  F_{-\nu0}|E_\perp(\hat{z}_e)|F_{00}\rangle\,, \nonumber \\
 & \!\!\!\!\!\!\!\!\!\!\!\!\!\!\!\!\!\!\!\!\!\!\!
 \langle\psi^{(0)}_{\nu0\mp}|\hat{V}_\perp|\psi^{(0)}_{00\pm}\rangle  =
  \langle\psi^{(0)}_{0\nu\mp}|\hat{V}_\perp|\psi^{(0)}_{00\pm}\rangle =
  \langle\psi^{(0)}_{\nu0\pm}|\hat{V}_\perp|\psi^{(0)}_{00\pm}\rangle =
  \langle\psi^{(0)}_{0\nu\pm}|\hat{V}_\perp|\psi^{(0)}_{00\pm}\rangle = 0\,,
\end{eqnarray}
where we have used 
\begin{equation}\label{F-relation1}
F_{nm}(z_e,z_h)=F'_{nm}(z_h,z_e)\,, % \qquad
% F_{00}(z_e,z_h)=F_{00}(z_h,z_e)\,,
\end{equation}
and that for a suitable choice of phases [cf.~Eq.~(\ref{envelope_F})]
\begin{equation}\label{F-relation2}
F_{nm}(z_e,z_h)=F_{mn}(z_h,z_e)\,.% F_{-\nu0}(z_e,z_h)=F_{0-\nu}(z_h,z_e)\,.
\end{equation}

Thus the contribution from $|\psi^{(0)}_{\nu0\pm}\rangle$
($|\psi^{(0)}_{0\nu\pm}\rangle$) corresponding to $E_{12}$
($E_{21}$) vanishes identically and, to first order in
perturbation theory, we obtain for the admixture of
cross-polarised excitons to the states $|\psi_{00\pm}\rangle$
induced by $E_\perp$
\begin{eqnarray}\label{wave-approx}
\!\!\!\!\!\!\!\!\!\!\!\!\!\!\!\!\!\!\!\!\!\!\!\!\!\!\!\!\!\!\!\!\!\!\!
   |\psi_{00\pm}^{(1)}\rangle\approx - \xi\sum_l
  \left(|\psi^{(0)}_{-\nu0\mp,l}\rangle-|\psi^{(0)}_{0-\nu\mp,l}\rangle
    +\zeta|\psi^{(0)}_{-\nu0\pm,l}\rangle
    +\zeta|\psi^{(0)}_{0-\nu\pm,l}\rangle\right)\langle  
  F_{-\nu0,l}|E_\perp(\hat{z_e})|F_{00}\rangle\,,\nonumber\\
\end{eqnarray} 
%\end{widetext}
where $\xi\equiv e R /2\epsilon_\perp(E_{13}-E_{11})$,
$\zeta\equiv\hbar/3m_ev_FR$ and $l$ labels a complete set of
envelope functions for the $E_{13}$ and $E_{31}$ manifolds. Here
we have assumed for the latter that those having appreciable
overlap with $|F_{00}\rangle$ correspond to excitons with
energies well approximated by the lowest one ($E_{13}=E_{31}$)
and have neglected the splitting between the bonding and
antibonding manifolds.  Note that these manifolds are likely to
consist of delocalised electron-hole pair like states as is the
case with the $E_{33}$ band so that the analysis in section
\ref{subsec:Xconfinement} may not apply \cite{Araujo07}. However
our treatment of $E_\perp$ relies only on the validity of
Eqs.~(\ref{wavefunction}) and (\ref{F-relation2}) irrespective
of the specific form of the envelope functions $F_{-\nu0,l}$,
and on satisfying the weakness criterion $|\xi
E_\perp(z_0)|^2\ll1$.

The Hamiltonian describing the interaction between electrons and
low-frequency phonons has two distinct terms: (i) a deformation
potential contribution diagonal in sublattice space
corresponding to an energy shift of the Dirac point, 
\begin{equation}\label{HamiltonianDP1}
  \hat{H}_\mathrm{e-ph}^{(1)} = g_1 \!\left[ \hat{u}_{\varphi\varphi}
    (\hat{\bar{r}}) + \hat{u}_{zz} (\hat{\bar{r}})\right]\,,
\end{equation}
and (ii) a bond-length change contribution off-diagonal in
sublattice space \cite{Suzuura02,Mariani09} that in the Dirac
picture emulates a gauge field,
\begin{eqnarray}\label{HamiltonianDP2}
  \hat{H}_\mathrm{e-ph}^{(2)} = & \,\, g_2 \!\left\{\left( \cos\! 3
      \theta \,\hat{\sigma}_{x'} \hat{\tau}_3 - \sin\!3\theta
      \,\hat{\sigma}_{y'} \right)\left[ \hat{u}_{\varphi\varphi}
      (\hat{\bar{r}}) - \hat{u}_{zz} (\hat{\bar{r}})\right]
  \right. \nonumber \\
  & \left. - \left( \sin\! 3 \theta \,\hat{\sigma}_{x'}
    \hat{\tau}_3 + \cos\!3\theta \,\hat{\sigma}_{y'} \right)
  2\hat{u}_{\varphi z} (\hat{\bar{r}})\right\} \,.
\end{eqnarray}
Here $g_1\approx30\,$eV is the deformation potential constant,
$g_2\approx1.5\,$eV the off-diagonal electron-phonon coupling
constant \cite{Suzuura02,Mariani09}, and $\hat{x},\hat{y}$ only
act on the ``orbital'' part of the 1D Bloch functions
(cf.~Eq.~\ref{states}).  The single-particle Hamiltonians
(\ref{HamiltonianDP1}) and (\ref{HamiltonianDP2}) naturally lead
to the following excitonic two-particle Hamiltonian \footnote{As
  in Eq.~(\ref{wave-approx}), here we % use that the single
  % particle Hamiltonian is invariant under time reversal and
  assume that only contributions involving virtual transitions
  to single exciton states are relevant.}
\begin{equation}\label{HamiltonianX-ph}
  \hat{H}_\mathrm{X-ph}=\hat{\mathbb{P}}_e\hat{H}_\mathrm{e-ph}
  \hat{\mathbb{P}}_e\otimes \hat{\mathbb{I}}_h
    - \hat{\mathbb{I}}_e\otimes\hat{\mathbb{P}}_h\hat{H}_\mathrm{e-ph} 
  \hat{\mathbb{P}}_h
\end{equation}
where
$\hat{H}_\mathrm{e-ph}=\hat{H}_\mathrm{e-ph}^{(1)}+\hat{H}_\mathrm{e-ph}^{(2)}$
and $\hat{\mathbb{P}}_e$ ($\hat{\mathbb{P}}_h$) denotes the
projector on the conduction (valence) band.

Given that the relevant phonon modes have wavelengths, $2\pi/q$,
that are much larger than the CNT radius, $R$, one can describe
them using a continuum shell model \cite{Suzuura02,Yakobson96},
with $\hat{u}_{ij}(\hat{r})$ the corresponding Lagrangian strain
operator, and keeping only the lowest orders in $q R$
\cite{Suzuura02,Yakobson96} which amounts to the use of thin rod
elasticity (TRE) \cite{Wilson-Rae08}.  Within this
approximation, both compressional and flexural deformations have
the structure of a local stretching so that the strain
components satisfy $u_{\varphi z}=0$,
$u_{\varphi\varphi}=-\sigma u_{zz}$ with the latter given,
respectively, by
\begin{equation}\label{strain}
  u_{zz}=-R \cos\!\varphi \,\frac{\partial^2 \phi_f}{\partial z^2}\,
  \qquad \mathrm{and} \qquad \,u_{zz}=\frac{\partial
    \phi_c}{\partial z}
\,,
\end{equation}
where $\phi_{f}(z)$ [$\phi_{c}(z)$] is the flexural
[compressional] 1D field \cite{Wilson-Rae08} that satisfies the
Euler-Bernoulli [classical-wave] equation and $\sigma=0.2$ is a
Poisson ratio for the CNT \cite{Yakobson96}. Finally, we
consider the lowest order contributions in the external field
(zeroth order in $\phi_\mathrm{AB}$) to the interaction between
the lowest bright exciton states
$|\psi_{00\pm}\rangle\approx|\psi_{00\pm}^{(0)}\rangle +
|\psi_{00\pm}^{(1)}\rangle$ of the NTQD and low-frequency
phonons, i.e.
%\begin{widetext}
\begin{equation}\label{QD-ph1}
  \langle\psi_{00\pm}|\hat{H}_\mathrm{X-ph}|\psi_{00\pm}\rangle
  \approx
  \langle\psi_{00\pm}^{(0)}|\hat{H}_\mathrm{X-ph}|\psi_{00\pm}^{(0)}\rangle
  + 2\Re\!
  \left[\langle\psi_{00\pm}^{(0)}|\hat{H}_\mathrm{X-ph}|\psi_{00\pm}^{(1)}
    \rangle\right]\,.
\end{equation}
%\end{widetext}
Naturally, the axial angular momentum is preserved by coupling
to compressional phonons and modified by interactions with
flexural phonons, so that only the first or the second term on
the R.H.S.\ of Eq.~(\ref{QD-ph1}) contributes, respectively, to the
corresponding NTQD deformation potentials
[cf.~Eqs.~(\ref{HamiltonianDP1})-(\ref{strain})].  Then, Eqs.~(\ref{states}),
(\ref{wavefunction}), (\ref{steps}), (\ref{eigen}), % (\ref{wave-approx}),
% (\ref{HamiltonianDP1}), (\ref{HamiltonianDP2}),
% (\ref{HamiltonianX-ph}), (\ref{strain}), (\ref{QD-ph1})
(\ref{F-relation1})--(\ref{QD-ph1}) and
\begin{equation}
 \nu\!=\!\pm1\quad\Rightarrow\quad \int_0^{2\pi}
 \!\frac{\rmd\varphi}{2\pi} \,e^{\pm 
   i\nu\varphi}\!\cos\!\varphi=\frac{1}{2}\,, 
\end{equation}
allow us to obtain
\begin{eqnarray}\label{HamiltonianQD-ph} 
\!\!\!\!\!\!\!\!\!\!\!\!\!\!\!\!\!\!\!\!\!\!\!\!\!\!\!
  \langle\psi_{00\pm}|\hat{H}_\mathrm{X-ph}|\psi_{00\pm}\rangle
  \approx \,& 2\nu g_2 (1+\sigma) \cos\!3\theta\,\langle
  F_{00}|\frac{\partial \hat{\phi}_c}{\partial
    z}(\hat{z}_e)|F_{00}\rangle +2\xi R \left[g_1(1-\sigma) \right.\nonumber\\
  & \left. + \,\nu \zeta g_2 (1+\sigma)
    \cos\!3\theta\right]\langle
  F_{00}|\frac{\partial^2\hat{\phi}_f}{\partial z^2}(\hat{z}_e)
  E_\perp(\hat{z}_e)|F_{00}\rangle
\end{eqnarray}
which is independent of the excitonic wavefunction's symmetry (bonding
vs. antibonding). Here we have exploited the completeness of the basis
$\{|F_{-\nu0,l}\rangle\}$. The ``off-diagonal'' coupling to flexural
phonons, i.e.~the term proportional to $\zeta$, constitutes a
negligible correction given that $\zeta=\alpha_Fc
a_B/3v_FR\lesssim0.1$ (for stable CNT radii \cite{Yakobson96}) and
$g_2/g_1\approx 0.05$.%Nano Lett., 2008, 8 (2), pp 459–462

To recapitulate, in the present section we have used an envelope
function approximation based on the graphene zone-folded description
% of a CNT
to analyse how weak inhomogeneous electrostatic fields affect the
axially-polarised direct excitons of a wide-gap semiconducting CNT. We
have shown that such perturbative fields can induce quantum
confinement of the CM motion dominated by the axial component and
characterised by a level spacing in the milli-electronvolt range, and
have derived the effective deformation potential (DP) for the
lowest-energy direct-exciton states of the resulting optically-active
quantum dot (NTQD). This approximation for the excitonic DP in the
presence of applied electrostatic fields, given by
Eq.~(\ref{HamiltonianQD-ph}), constitutes a central result of this
paper. %article
It relies on: (i) the aforementioned envelope function approximation,
(ii) the existence of a weakly confined Wannier-exciton state
$|\psi_{00}\rangle$ ---which requires the adiabatic conditions on
$E_\parallel(z)$ detailed in Eq.~(\ref{adiabatic}), (iii) neglecting
the dispersion of the relevant cross-polarised excitonic manifold and
(iv) the weakness criterion on the transverse field $|\xi
E_\perp(z_0)|^2\ll1$ \footnote{Note that $1/|\xi|$ is comparable to
  the transverse critical field for exciton dissociation ---
  $1/|\xi|\sim10^4\,\mathrm{V}\mu^{-1}$ for relevant nanotube parameters
  (cf.~Sec.~\ref{subsec:laser-cool}).}. This effective excitonic DP
involves coupling to both compressional (i.e.~stretching) and flexural
phonons. The latter is linear in the transverse field $E_\perp$ which
induces a hybridisation with the cross-polarised excitons.
% leading to an interaction with the flexural phonons of the CNT that
% is linear in $E_\perp$.
Interestingly within our envelope function approximation (i), the 1D
subband Bloch functions alternate between the two orthogonal
eigenstates of the sublattice pseudospin $\hat{\sigma}_{x'}$ as the
corresponding eigenenergy is increased ---cf.~Eqs.~(\ref{states}) and
(\ref{energies}).  This leads to the selection rules embodied in
Eqs.~(\ref{matrix-x2}) and (\ref{matrix-x2prime}), which imply that
the contribution from the $E_{12}$ and $E_{21}$ excitons vanishes.

\section{Exciton-assisted optomechanics}\label{sec:Xoptomechanics}

In this section we consider the scenario relevant for
optomechanics where axial phonon-confinement due to the finite
suspended-length of the CNT plays a leading role. Whence the
flexural and compressional phonon branches of the CNT are
characterised by sharp resonances with a free spectral range
comparable to the optical linewidth of the zero phonon line
(ZPL) of the transition associated to the state
$|\psi_{00-}\rangle$ (analogous considerations apply to
$|\psi_{00+}\rangle$). We focus on laser excitation of the
latter near resonant with the ensuing lowest-frequency flexural
phonon sidebands.  More precisely, we consider a bridge geometry
(cf.~Fig.~\ref{fig:tips}) with a length ($L\sim100\,$nm) short
enough that the relative strength of these phonon sidebands is
weak, but long enough that phonon-radiation losses
(i.e.~clamping losses) are negligible \cite{Wilson-Rae08}.
Hereafter, we follow Ref.~\cite{Wilson-Rae04} in which an
analogous scenario was analysed in a semiconductor
heterostructure.

% We first discuss in Sec.~\ref{subsec:confinement} the
% consequences of the phonon axial confinement
%\subsection{Phonon-confinement}\label{subsec:confinement}

To this effect we use the formalism developed in
Ref.~\cite{Wilson-Rae08} and adopt a resonator-bath
representation with the resonator mode (annihilation operator
$b_0$, angular frequency $\omega_0$, and quality factor $Q$) 
% \footnote{For high-stress flexural CNT resonators in the GHz
%   range $Q$-values approaching $10^5$ have recently been
%   measured \cite{Huttel09}. Here we consider a stress-free
%   resonator, but the additional fact that dissipation in these
%   systems is nonlinear \cite{Eichler11} implies that the
%   \emph{linear} $Q$-values relevant in our context could still
%   be high --- note that the distinction between high-stress and
%   low-stress regimes depends on the length.}
corresponding to the fundamental in-plane flexural resonance
(that we intend to manipulate and laser cool to the ground
state) and the bath including the other nanotube vibrational
resonances coupled to the 3D substrate that supports the
CNT. Thus, we use for the effective field operators
$\hat{\phi}_f$, $\hat{\phi}_c$ in Eq.~(\ref{HamiltonianQD-ph})
the following resonator-bath mode decomposition
\begin{eqnarray}\label{decomposition}
%\!\!\!\!\!\!\!
 & \frac{\partial^2\hat{\phi}_f}{\partial
  z^2}(\hat{z}_e) \to \frac{\partial^2 \phi_0}{\partial
  z^2}(\hat{z}_e)\sqrt{\frac{\hbar}{2\mu\omega_0}}\,b_0%+b_0^{\dag})%\nonumber\\
% &
+\sum_q \frac{\partial^2 u_{x,q}}{\partial
  z^2}(\hat{z}_e)\sqrt{\frac{\hbar}{2\rho_s\omega(q)}}\,b_q+
\mathrm{H.c.}%+b_q^{\dag})  
\,, \nonumber\\ 
%\!\!\!\!\!\!\!
& \frac{\partial\hat{\phi}_c}{\partial z}(\hat{z}_e)\to   \sum_q
\frac{\partial u_{z,q}}{\partial
  z}(\hat{z}_e)\sqrt{\frac{\hbar}{2\rho_s\omega(q)}}\, b_q+
\mathrm{H.c.}%b_q^{\dag})\,. 
\end{eqnarray}
Here $\phi_0(z)$ is the normalised resonator 1D eigenmode and
$\mu$ is the linear mass density of the CNT; while $u_{x,q}(z)$
[$u_{z,q}(z)$] with $z\in\,$CNT is the $x$ [$z$] component of
the centre of mass displacement of the CNT cross section at $z$
(cf.~Fig.~\ref{fig:tips}) for the bath mode (annihilation
operator $b_q$) corresponding to the scattering eigenmode $q$
\cite{Wilson-Rae08}, and $\rho_s$ is the mass density of the
substrate. Subsequently, substituting Eq.~(\ref{decomposition})
into Eq.~(\ref{HamiltonianQD-ph}) we obtain
\begin{equation}\label{HamiltonianQD-ph2}
  \!\!\!\!\!\!\!\!\!\!\!\!\!\!
  \!\!\!\!\!\!\!\!\!\!\!\!\!\!|\psi_{00\pm}\rangle\langle\psi_{00\pm}| 
  \hat{H}_\mathrm{X-ph} |\psi_{00\pm}\rangle 
  \langle\psi_{00\pm}|= \hbar\left(\eta\omega_0 b_0 + \sum_q
    \lambda_q b_q + \mathrm{H.c.} 
  \right)|\psi_{00\pm}\rangle\langle\psi_{00\pm}|  %+ \mathrm{H.c.}
\end{equation}
with
\begin{eqnarray}\label{lambda}
  %\!\!\!\!\!\!\!\!\!\!\!\!\!\!\!\!\!\!\!\!\!\!\!\!\!\!\!
  \lambda_q\approx \,& \sqrt{\frac{\hbar}{2\rho_s\omega(q)}}\left\{2\nu
    g_2 (1+\sigma) \cos\!3\theta\,\langle 
    F_{00}|\frac{\partial u_{z,q}}{\partial
      z}(\hat{z}_e)|F_{00}\rangle 
    \right.\nonumber\\ 
  & \left. % + \,\nu \zeta g_2 (1+\sigma)
      % \cos\!3\theta\right]
    +\,2\xi R g_1(1-\sigma)\langle
    F_{00}|\frac{\partial^2u_{x,q}}{\partial z^2}(\hat{z}_e)
    E_\perp(\hat{z}_e)|F_{00}\rangle\right\}
\end{eqnarray}
and
\begin{equation}\label{eta}
  \eta\approx 2^{3/4}(1-\sigma)
  \frac{g_1 \sigma_G^{1/4}\xi\mathcal{E}_\perp}{R E_G^{3/4}
    \left(q_0 L\right)}\sqrt{\frac{L}{\pi\hbar}}\,,
\end{equation}
where we have introduced the effective field
$\mathcal{E}_\perp\equiv(\sqrt{L}/q_0^2)\langle
F_{00}|\frac{\partial^2\phi_0}{\partial
  z^2}(\hat{z}_e)E_\perp(\hat{z}_e)|F_{00}\rangle$, set
$\zeta\to0$ and used $\omega_0=\tilde{c}_cRq_0^2/\sqrt{2}$.
Here $q_0\approx 4.73/L$ is the fundamental mode's
Euler-Bernoulli wavevector for clamped-clamped boundary
conditions, and $E_G=340\,$Nm$^{-1}$ and
$\sigma_G=7.6\times10^{-7}\,$kgm$^{-2}$ are, respectively, the
2D Young modulus \cite{Lee08} and mass density of graphene
($a=0.246\,$nm \cite{Barros06}). The parameter $\eta$ is the
``modulation parameter'' associated to the phonon sidebands that
characterises the strength of the exciton-resonator coupling
with $e^{-\eta^2/2}$ corresponding to the Franck-Condon factor
($\eta^2$ is also known as the Huang Rhys parameter). We will
focus mainly on regimes such that $\eta\ll1$. Note that the
perturbative nature of $E_\perp$ underpinning Eq.~(\ref{eta})
implies $\xi\mathcal{E}_\perp\ll1$.

Then from Eq.~(\ref{HamiltonianQD-ph2}), adopting Pauli matrix
notation for the optical pseudospin ($\sigma_z=1$ corresponds to
$|\psi_{00-}\rangle$ and $\sigma_z=-1$ to the empty neutral
NTQD) and a frame rotating at the laser frequency $\omega_L$,
and applying the shift
\begin{equation}\label{shift}
  \hat{O}'=e^{-\frac{\eta}{2}
    (b_0-b_0^{\dag})-\sum_q\frac{\lambda_q}{2\omega_q}
    (b_q-b_q^{\dag})}\,\hat{O}\,e^{\frac{\eta}{2}
    (b_0-b_0^{\dag})+\sum_q\frac{\lambda_q}{2\omega_q}
    (b_q-b_q^{\dag})}
\end{equation}
to the phonon modes, we obtain the following Hamiltonian for the
laser driven NTQD coupled to the resonator mode ($\hbar=1$)
\begin{equation}\label{H_sys_original}
  H'_\mathrm{sys}= -\frac{\delta}{2} \sigma_z +\frac{\Omega}{2}\sigma_x +
  \frac{\eta\omega_0}{2}\sigma_z\left(b_0^{\vphantom{\dag}}+b_0^{\dag}\right)+
  \omega_0 b_0^\dag b_0^{\vphantom{\dag}}\,. 
\end{equation}
Here $\delta$ is the laser detuning from the ZPL and $\Omega$
the Rabi frequency.

To analyse the dynamics that arises from the interplay between
$H'_\mathrm{sys}$ and the couplings to the phonon bath and to
the radiation field (annihilation operators $a_k$ and couplings
$g_k$), we perform the following canonical transformation
\cite{Wilson-Rae04}
\begin{equation}\label{polaronic}
  \hat{O}'=e^{-\frac{\eta}{2}\sigma_z
    \left(b_0^{\vphantom{\dag}}-b_0^{\dag}\right)}\,\hat{O}\,e^{\frac{\eta}{2}\sigma_z
    \left(b_0^{\vphantom{\dag}}-b_0^{\dag}\right)}
\end{equation}
to a ``polaronic'' representation. The resulting complete
Hamiltonian consists of three contributions,
$H=H_\mathrm{sys}+H_\mathrm{int}+H_B$, given by ($\hbar=1$)
\begin{eqnarray}
  H_\mathrm{sys}= &-\frac{\delta}{2} \sigma_z + \frac{\Omega}{2}
  \left(\sigma_+ B^\dag+ \sigma_- B^{\vphantom{\dag}}
  \right)+ \omega_0 b_0^\dag b_0^{\vphantom{\dag}} \label{spinboson_sys} \\
  H_\mathrm{int}= & \sum_k g_k \sigma_+ B^\dag a_k^{\vphantom{\dag}} +
  \sum_q\zeta_q b_0^\dag b_q^{\vphantom{\dag}} + \frac{\sigma_z}{2}
  \sum_q \lambda_q b_q +\mathrm{H.c.} \label{spinboson_int} \\
  H_B= & \sum_q\omega_q b_q^\dag b_q^{\vphantom{\dag}} +\sum_k
  (\omega_k-\omega_L) a_k^\dag a_k^{\vphantom{\dag}}\,,\label{spinboson_bath}
\end{eqnarray}
where $B\equiv e^{\eta (b_0-b_0^{\dag})}$. The couplings
$\zeta_q$ and $\lambda_q$ to the phonon bath lead, respectively,
to the resonator-mode's phonon-radiation losses and to pure
dephasing of the NTQD induced by the deformation potential
[cf.~Eqs.~(\ref{HamiltonianQD-ph2}) and (\ref{lambda})].  To
assess their role we note that: (i) $q R\ll1$ and (ii) the
reflection symmetry of the whole structure with respect to the
$y-z$ plane, imply that the $q$ can be effectively identified
with the low-frequency CNT phonon branches so that the couplings
$\zeta_q$ involve only modes with a flexural character
\cite{Wilson-Rae08}. The RWA for the latter is justified given
$Q\gg1$ and $\eta\ll1$. These conditions and the anharmonicity
of the flexural spectrum also imply that the effects of the
$\lambda_q$ associated to the flexural branch can be neglected
\cite{phdthesis}.

In turn, the pure dephasing induced by the compressional
(stretching) phonons is determined by the low-frequency behaviour
of the spectral density
\begin{equation}
J_c(\omega)=\pi\sum_q|\lambda_q|^2\delta(\omega-\omega_q)
\end{equation}
with $q\in$ compressional branch. The latter can be obtained
from Eq.~(\ref{lambda}), building upon the scattering formalism
previously developed (cf.~Eq.~(34) in Ref.~\cite{Wilson-Rae08})
by incorporating now the leading correction in $q R$ to the
zeroth order 1D-3D reflection amplitude at a single junction
between the resonator and the support (clamping point) and
considering the ``single support'' case in which both clamping
points connect to the same elastic half-space so that
interference effects are crucial --- as relevant to our scenario
where the gap below the CNT satisfies $\lesssim L$
(cf.~Fig.~\ref{fig:tips}). This procedure which will be detailed
elsewhere yields a superohmic phonon spectral density
\cite{Leggett87},
\begin{equation}\label{Jc_con}
  J_c(\omega)\approx2\pi\alpha_\mathrm{con}\frac{\omega^3}{\omega_*^2}
  \ \quad \mathrm{for} \ \quad\omega\ll \omega_*\,,
\end{equation}
with
\begin{equation}\label{alphacon}
  \alpha_\mathrm{con}\sim\frac{\alpha}{\pi
    Q_c}\qquad\mathrm{and}\qquad\omega_*\sim\frac{c_{\mathrm{SAW}}}{L}\,,
\end{equation}
where 
\begin{equation}
Q_c=\frac{0.22}{\pi^2} \sqrt{\frac{\sigma_G}{\rho_s}
  \left(\frac{E_s}{E_G}\right)^3 }\frac{L^2}{R}  
\end{equation}
is the clamping-loss limited $Q$-value of the fundamental
compressional mode for the ``double support'' case
\cite{Wilson-Rae08} ($Q_c\sim 50$ for $L\sim100\,$nm), and
$\alpha$ is the dimensionless dissipation parameter
\cite{Leggett87} for the corresponding Ohmic (1D continuum)
result (note that the size of the QD-molecule generated by the
tip electrodes is set by $z_0$)
\begin{equation}\label{Jc_infty}
  J_c(\omega)=2\pi\alpha\omega \ \quad \mathrm{for} \ \quad
  \omega\ll \omega_U\sim\tilde{c}_c/z_0
\end{equation}
which is valid in the limit $L\to\infty$ (keeping $\omega$
fixed). Here $c_{\mathrm{SAW}}$ and $E_s$ are, respectively, the phase
velocity of surface acoustic waves and Young modulus of the substrate
which is assumed to have a Poisson ratio $\nu_s=1/3$.

The dissipation parameter $\alpha$ plays a crucial role in
determining the different phases of the corresponding spin-boson
model [Eqs.~(\ref{spinboson_sys})-(\ref{spinboson_bath}) with
$\eta=g_k=\zeta_q=0$] in particular whether excitonic Rabi
oscillations are possible for $\Gamma\ll\Omega<\omega_U$
\cite{Leggett87}. Using the results of
Sec.~\ref{subsec:X-phonon} it is straightforward to obtain
\begin{equation}\label{alpha}
  \alpha = \frac{g_2^2\sqrt{\sigma_G}(1+\sigma)^2}{2\pi^2\hbar R E_G^{3/2}}
  \cos^2 3\theta\,
\end{equation}
which has a marked dependence on the chirality angle
$\theta$. We note that for stable CNT radii Eq.~(\ref{alpha})
always yields $\alpha<0.02$. In turn, a straightforward analysis
of the linear absorption spectrum (i.e.~for $\Omega\ll\Gamma$)
that results from this weak-coupling continuum Ohmic model
\cite{Leggett87} (in the relevant low temperature regime
$k_\mathrm{B}T/\hbar\ll \omega_U$) implies a pure-dephasing
contribution to the linewidth given by $\gamma_\mathrm{ph}=2\pi
\alpha k_\mathrm{B}T/\hbar$ and an asymmetric power law tail
$\sim \delta^{2\alpha-1}$ which is only observable for
$\max\{\Gamma,\gamma_{\mathrm{ph}}\}\ll 2^{-1/2\alpha}$. The
confined scenario embodied by Eq.~(\ref{Jc_con}) leads to a
favourable suppression of this phonon-induced decoherence which
would otherwise preclude achieving the desired resolved-sideband
condition for realistic parameters
(cf.~Sec.~\ref{subsec:laser-cool}). Thus, this feature
constitutes a cornerstone for the scheme proposed here. More
precisely, the superohmic behaviour with $n\geq3$ and
$\alpha_\mathrm{con}\ll1$ implies that for relevant ratios
$\Gamma/\omega_0$ and $\Omega\lesssim\omega_0$
(cf.~\ref{subsec:laser-cool}), the impact of these ``background
modes'' on optomechanical effects involving the resonator mode
is completely negligible \cite{phdthesis} --- henceforth we take
$\lambda_q\to0$.

In the following subsections we show how the resulting
optomechanical system can be used for ground-state cooling of
the fundamental mode (Sec.~\ref{subsec:laser-cool}), consider
the main practical limits to the strength of the optomechanical
coupling that can be achieved
(Sec.~\ref{subsec:implementation}), and discuss the prospects
for reaching the corresponding strong and ultra-strong coupling
regimes (Sec.~\ref{subsec:strong-coupling}).

\subsection{Laser-cooling}\label{subsec:laser-cool}

To analyse laser cooling of the resonator mode we derive, first,
a reduced master equation for the NTQD-resonator system and,
subsequently, a rate equation for its populations. For the
relevant regimes all environmental couplings can be treated
within the Born-Markov approximation and, after eliminating the
bath phonon modes and the radiation field, the Liouvillian of
the driven NTQD coupled to the resonator mode reads
\begin{eqnarray}\label{master}
  \dot{\rho}= & -i\left[H_\mathrm{sys},\rho\right]+
  \textstyle{\frac{\Gamma}{2}}\!\left(2\sigma_{-} B\rho
    B^{\dagger}\sigma_{+}-\rho
    \sigma_{+}\sigma_{-}-\sigma_{+}\sigma_{-}\rho\right)\nonumber\\
  & + \omega_{0}
  \textstyle{\frac{n\left(\omega_{0}\right)}{2Q}}\!\left(2
    b_0^{\dagger}\rho b_0^{\vphantom\dagger} -
    b_0^{\vphantom\dagger}b_0^{\dagger}\rho - \rho
    b_0^{\vphantom\dagger}b_0^{\dagger} \right)  \nonumber\\
  & +\omega_{0}\textstyle{\frac{n\left(\omega_{0}\right)
      +1}{2Q}}\!\left(2 b_0^{\vphantom\dagger}\rho b_0^{\dagger}
    - b_0^{\dagger}b_0^{\vphantom\dagger}\rho - \rho
    b_0^{\dagger}b_0^{\vphantom\dagger}\right)\,,
\end{eqnarray}
where $n(\omega_0)$ is the thermal equilibrium occupancy at the
ambient temperature,
\begin{equation}
  \Gamma=2\pi\sum_k|g_k|^2\delta(\omega_L-\delta-\omega_k) 
\quad\mathrm{and}\quad \frac{\omega_0}{Q}=
2\pi\sum_q|\zeta_q|^2\delta(\omega_0-\omega_q)\,.
\end{equation}
Other relevant sources of dissipation beyond those considered in
Hamiltonian Eqs.~(\ref{spinboson_sys})-(\ref{spinboson_bath});
namely, internal losses of the CNT resonator %\cite{DeMartino09}
and nonradiative recombination of the exciton \cite{Hogele08}
can be incorporated, respectively, by simply adopting modified
values of $Q$ and $\Gamma$. We note that the nonradiative
recombination rate for an ultra-clean CNT dark exciton has not
been measured so far.

In analogy with the Lamb-Dicke limit (LD approximation) we
expand up to second order the translation operators $B$ and
adiabatically eliminate the NTQD to obtain a rate equation for
the populations of the resonator mode's Fock states
\cite{Wilson-Rae08NJP}. The latter reads \cite{Wilson-Rae04}
\begin{eqnarray}\label{rate:equ}
\!\!\!\!\!\!\!\!\!\!\!\!\!\!\!\!
\dot{P}_n = & \left[ \eta^2 A_+ + \omega_0 \textstyle{\frac{n \left(
   \omega_0 \right)}{Q}} \right] \left[n P_{n-1} - \left( n+1 \right)
   P_ n \right] + \left[ \eta^2 A_- + \omega_0
   \textstyle{\frac{n \left( \omega_0 \right) + 1}{Q}} \right]\nonumber\\
   & \times\left[\left( n+1 \right) P_{n+1} - n P_n \right]
\end{eqnarray}
with the cooling ($\eta^2A_-$) and heating ($\eta^2A_+$)
transition rates per quanta determined by 
% A_\mp=(2 \Gamma (4 (\mp + \delta)^2 + \Gamma^2) \Omega^2 (2 +
% 2 \Gamma^2 + \Omega^2))/((4 \delta^2 + \Gamma^2 + 2 \Omega^2)
% ((1 + \Gamma^2) (4 (-1 + \delta)^2 + \Gamma^2) (4 (1 +
% \delta)^2 + \Gamma^2) + 4 (4 \delta^2 (2 + \Gamma^2) + (-2 +
% \Gamma^2) (4 + \Gamma^2)) \Omega^2 + 4 (4 + \Gamma^2)
% \Omega^4))
$A_\mp=\Gamma 
    [(\delta\mp\omega_0)^2 + 
    \Gamma^2/4] (\omega_0^2 + 
    \Gamma^2 + \Omega^2/2)\Omega^2/D$, where
\begin{eqnarray}\label{rates}
   \!\!\!\!\!\!\!\!\!\!\!\!\!\!\!\!\!\!\!\!\!\!\!
  D\equiv &
  \!\!\!\!\!\!\!\!\! 4\left(\delta^2 + \frac{\Gamma^2}{4}
    +\frac{\Omega^2}{2}\right) 
  \left\{\left(\omega_0^2 + \Gamma^2\right)
    \left[\left(\delta-\omega_0 \right)^2
      +\frac{\Gamma^2}{4}\right] \left[\left(\delta+\omega_0 
        \right)^2 + \frac{\Gamma^2}{4}\right]
  \right. \nonumber\\ & \!\!\!\!\!\!\!\!\! \left.+ 
    \left[\delta^2 \left(2\omega_0^2 + 
        \Gamma^2\right) 
      - \left(2\omega_0^2 -  
        \Gamma^2\right) \left(\omega_0^2 +
        \frac{\Gamma^2}{4}\right)\right] \Omega^2 + 
    \left(\omega_0^2 + \frac{\Gamma^2}{4}\right) 
    \Omega^4\right\}\,.
\end{eqnarray}
We note that in this system, in stark contrast to atomic laser
cooling, $A_\mp\to0$ in the limit $\Omega\to\infty$ --- in the
polaronic representation this is associated to quantum
interference effects between multiple
paths\cite{Wilson-Rae04}. Thus the steady state occupancy for
$Q\to\infty$, i.e.~the quantum backaction limit, reduces to
\begin{equation}
  \frac{A_+}{A_--A_+}=-\frac{\left(\delta+\omega_0 
        \right)^2 + \frac{\Gamma^2}{4}}{4\delta}\,.
\end{equation}
Remarkably, this result is independent of the Rabi frequency and
coincides with the expression valid for the optomechanical
cavity-assisted backaction cooling \cite{Wilson-Rae07,Marquardt07}
(Eq.~(6) in Ref.~\cite{Wilson-Rae07}) with the cavity decay rate
$1/\tau$ replaced by the spontaneous emission rate which for the
optimal detuning $\delta=-\sqrt{\omega_0^2+\Gamma^2/4}$ yields the
fundamental limit $(\sqrt{1+(\Gamma/2\omega_0)^2}-1)/2$. This
coincidence can be better understood by performing the adiabatic
elimination in the ``original'' representation corresponding to
Eq.~(\ref{H_sys_original}) --- a thorough analysis using this
representation has been performed by Jaehne et al.~\cite{Jaehne08}.

\begin{figure}
  \fig{\linewidth}{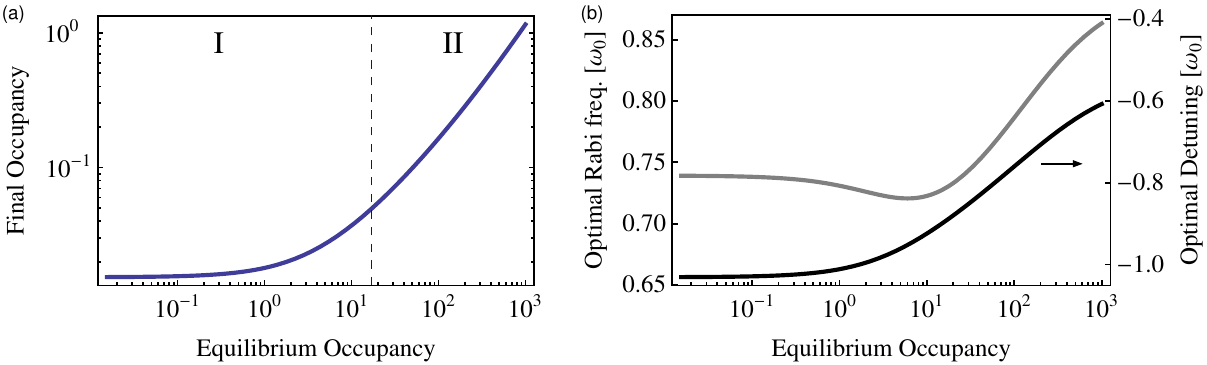}
  \caption{(a) Optimal steady state (final) occupancy ($n_f$)
    v.s.~equilibrium (thermal) occupancy $n(\omega_0)$ for
    $\Gamma/\omega_{0} = 1/2$ and $\eta^{2} Q =1.1\times\!10^3$. The
    optimum results from minimising $n_{f}$ with respect to the laser
    parameters. For $n(\omega_0) \leq(\frac{\Gamma}{4
      \omega_{0}})^{2}$ laser cooling is not operational. For
    $n(\omega_0) \geq(\frac{\Gamma}{4 \omega_{0}})^{2}$ there are two
    regimes (dotted line): I) when $n(\omega_0) \ll\eta^{2} Q
    (\frac{\Gamma}{4 \omega_{0}})^{2} $, $n_{f}$ is determined by the
    quantum backaction limit $\approx (\frac{\Gamma}{4
      \omega_{0}})^{2}$ and II) when $n(\omega_0)$ is large
    [$n(\omega_0) \gg\eta^{2} Q (\frac{\Gamma}{4 \omega_{0}})^{2} $]
    thermal noise dominates and the optimum becomes linear in the
    ambient temperature. (b) Rabi frequency (gray) and laser detuning
    (black) for which $n_f$ is minimised (in units of $\omega_0$).
  \label{fig:finalT}}
\end{figure}

Figure~\ref{fig:finalT} shows the resulting temperature
dependence of the optimal steady state occupancy ($n_f$) that
follows from Eq.~(\ref{rate:equ}), i.e.
\begin{equation}
\langle b_0^{\dagger}b_0^{\vphantom\dagger}\rangle_\mathrm{SS}=\frac{n(\omega_0)
+ \eta^2 Q A_+/\omega_0}{1 + \eta^2 Q (A_- - A_+)/\omega_0}%\,,
\end{equation}
minimised with respect to the laser detuning $\delta$ and Rabi
frequency $\Omega$, for typical parameters. These correspond for
instance to $\!\Gamma=\omega_0/2$, $Q=1.5\times\!10^{5}$ and
$\eta=0.086$. The latter can be realised for example with a $(9,4)$
nanotube [cf.~Eq.~(\ref{eta})] \footnote{Since to our knowledge the
  optical transition energy $E_{13}$ has so far not been determined,
  we take as a rough estimate for the energy difference in $\xi$,
  $E_{13}-E_{11}\sim E_{12}+ 2\hbar v_F/3R$ which corresponds to
  assuming that the correlation energy for $E_{13}$ is the same as for
  the cross-polarised excitons $E_{12}$, and use for the latter the
  value reported in Ref.~\cite{Kilina08} --- this procedure yields
  $1.3\,$eV.}  ($R=0.45$nm) of length $L=120\,$nm
($\omega_0/2\pi=1.67\,$GHz) and a transverse effective field
$\mathcal{E}_\perp=36.8\,$V$\mu^{-1}$ (corresponding to the field
configuration in Fig.~\ref{fig:tips}). For this case an ambient
temperature $T=4.2\,$K [$n(\omega_0)=52$] allows cooling to a final
(steady state) occupancy $n_f=0.10$. In the high temperature regime
(II) the cooling factor is constant and straightforward arguments
imply that its optimum is approximately bounded by $\eta^2 Q \lesssim
n(\omega_0)/n_f \lesssim \eta Q$ \cite{Wilson-Rae04,Jaehne08}. We note
that for the parameters considered here (cf.~Fig.~\ref{fig:tips}) in
this linear regime the initial effective modulation parameter is no
longer small, i.e.~$\eta^2 [n(\omega_0)+1]\gtrsim1$ and we find for
the optimum
$A_-|_\mathrm{opt}\sim(A_--A_+)|_\mathrm{opt}\sim\omega_0$, so that
initially $\eta^2 [\langle
b_0^{\dagger}b_0^{\vphantom\dagger}\rangle+1]A_-\gtrsim\Gamma$ and
Eq.~(\ref{rates}) fails to describe the complete cooling process which
will suffer a slowing down. Nonetheless for these parameters $\eta^2
[n_f+1]A_-|_\mathrm{opt}/\Gamma<0.03$ for all $n(\omega_0)<10^3$, and
the LD approximation remains valid for the steady state --- these
aspects have been analysed in detail by Rabl \cite{Rabl10}.

\subsection{Implementation}\label{subsec:implementation}

The choice of parameters in section \ref{subsec:laser-cool}
takes into account the restrictions imposed on the applied
electric fields by the need to avoid: deleterious static
deflections due to gradient forces \cite{Unterreithmeier09},
field emission, and dielectric breakdown
(cf.~Fig.~\ref{fig:tips}). For the nanotube the latter is taken
care of by the aforementioned restrictions imposed on the fields
$E_\perp(z_0)\sim E_\perp(0)$ and $E_\parallel(z_0)$ so as to
avoid dissociation of the $E_{11}$ exciton
(cf.~Sec.~\ref{sec:Xmodel}), while the suspended tip geometry of
the electrodes (cf.~Fig.~\ref{fig:tips}) allows for a strong
suppression of the field at the dielectric substrate with
respect to the maximum field. In fact, static deflections pose
the most stringent limitation when trying to maximise the
exciton-resonator coupling $\eta$, given that they scale as
\begin{equation}\label{deflection}
  \Delta x\sim\frac{\left(\alpha_\parallel+\alpha_\perp\right)}{\pi^5 E_G}
  \left(\frac{L}{R}\right)^3E_\perp(0)E_\parallel(z_0)\, 
\end{equation}
---here the prefactor is of order unity provided that the transverse
dimension of the electrodes is comparable to or smaller than $z_0<L$,
and we have used the reflection symmetries which imply that the total
electrostatic force on the nanotube scales as
$2(\alpha_\parallel+\alpha_\perp)E_\perp(0)E_\parallel(z_0)$. Note
that due to the Duffing geometric nonlinearity associated to a bridge
geometry this static deflection would result mainly in an effective
stiffening of the resonator for $\Delta x\gtrsim R$. For the
parameters considered in Sec.~\ref{subsec:laser-cool}, a detailed
calculation based on the FEM electric fields
(cf.~Sec.~\ref{subsec:Xconfinement}), the Euler-Bernoulli theory for
the CNT flexural response, and static polarisabilities
[$\alpha_\perp=13(4\pi\epsilon_0$\AA$^2)$, $\alpha_\parallel=75
(4\pi\epsilon_0$\AA$^2)$] estimated from
Refs.~\cite{Benedict95,Fagan07b} yields $\Delta x=0.013\,$nm (note
$R=0.45$nm). Under these conditions we find that the modification of
the resonator's frequency $\omega_0\to\tilde{\omega}_0$ induced by the
electric fields and the effect of the Duffing geometric nonlinearity
(which lead, respectively, to softening and stiffening) is completely
negligible ($\sim0.1\%$). In general, this modification also results
in a correction to the NTQD modulation parameter given by $\eta \to
(\omega_0/\tilde{\omega}_0)^{3/2}\eta$.

The proposed device is compatible with a confocal setup but
places stringent requirements on the nanotube positioning needed
to enforce the desired reflection symmetries. We note that the
latter are not essential to our scheme but just a convenient
means to enable a wider parameter range. If they are relaxed,
while it is still possible to tune independently $\Delta E$ and
$\eta$ via the two voltages $V_1$ and $V_2$, $|V_2-V_1|\gg
|V_2+V_1|$ can now result in regions where $E_\perp\sim
E_\parallel$ which reduces the maximum feasible $E_\perp$ given
that $E_\parallel$ should not exceed the critical field for the
breakdown of an $E_{11}$ exciton. Nonetheless, relevant
couplings
% $\eta\sim 0.04$
could still be achieved [cf.~Eq.~(\ref{eta})]. In turn for such
an asymmetric arrangement the NTQD may also couple to the
out-of-plane flexural resonances but this has no impact on the
validity of our effective model
[Eqs.~(\ref{spinboson_sys})-(\ref{spinboson_bath}) with
$\lambda_q=0$ for $q\in$ flexural branch] given that the
electrostatic fields and the mechanical coupling to the
substrate break the degeneracy between the two flexural
branches. A variant of this asymmetric scenario, albeit
relinquishing the independent tunability of $\Delta E$ and
$\eta$, is to use a single voltage configuration in which the
fields are applied with an STM tip. Another interesting option
is to effect the optical excitation of the nanotube using the
evanescent field of a tapered fibre which could allow for
instance to cancel the static deflection by using the optical
gradient forces induced by an additional off-resonant laser,
thereby lifting the restriction imposed on the length by
Eq.~(\ref{deflection}) \cite{Rips12}.

As the required fields are relatively large
($\sim10$V$\mu^{-1}$) and inhomogeneous, it seems natural to
also analyse the impact of voltage fluctuations arising from the
Johnson\-Nyquist noise of the electrodes and/or of the voltage
sources used. These induce a fluctuation of the effective
potential defining the NTQD [cf.~Eq.~(\ref{Veff})]
\begin{equation}\label{deltaVeff}
  \delta \hat{V}_\mathrm{eff}^{(V)}(\hat{z}_\mathrm{CM})=-2\alpha_X^{(00)}
  \frac{E_\parallel^2(\hat{z}_\mathrm{CM})}{V_1+V_2} \left(\delta
    \hat{V}_1+\delta \hat{V}_2\right) 
\end{equation} 
which given $k_BT<\Delta E$, only affects the exciton as a pure
dephasing contribution described by the interaction Hamiltonian $H_p=
\frac{1}{2}\sigma_z \hat{V}_p$ with $\hat{V}_p=\langle F_{00}|\delta
\hat{V}_\mathrm{eff}^{(V)}(\hat{z}_\mathrm{CM})|F_{00}\rangle$. Here
the white noise variables $\delta \hat{V}_1$ and $\delta \hat{V}_2$
are uncorrelated and determined by $\langle\delta
\hat{V}_{1/2}(t)\delta \hat{V}_{1/2}(0)\rangle=2R_Jk_BT\delta(t)$
where $R_J$ and $T$ are the relevant resistance and temperature ---for
estimating an upper bound we assume without loss of generality that
the noise is symmetric and dominated either by the electrodes or the
unfiltered sources. Then within the Markov approximation, these
voltage fluctuations result in an excess linewidth (FWHM)
$\gamma_p=\int_{-\infty}^{\infty}
\langle\hat{V}_p(t)\hat{V}_p(0)\rangle \rmd t/\hbar^2$. The latter can
be estimated using Eqs.~(\ref{deltaVeff}), (\ref{Xpolarizability}) and
$\langle F_{00}|E_\parallel^2(\hat{z}_\mathrm{CM})|F_{00}\rangle\sim
E_\parallel^2(z_0)$ leading to
\begin{equation}\label{gamma_p}
  \gamma_p\sim R_Jk_BT
  \left[\frac{32\pi\epsilon_0\epsilon_\parallel}{\hbar
      \left(V_1+V_2\right)}\right]^2 \sigma_{eh}^6E_\parallel^4\,.
\end{equation} 
Finally, for the envisaged parameters
(cf.~Sec.~\ref{subsec:Xconfinement} and Fig.~\ref{fig:tips}),
$R_J<1\mathsf{\Omega}$ and $T<300\,$K, Eq.~(\ref{gamma_p})
yields $\gamma_p/2\pi\lesssim 200\,$kHz which is negligible
compared to the values for $\Gamma\gtrsim100\,$MHz considered
here (cf.~Secs.~\ref{subsec:laser-cool} and
\ref{subsec:strong-coupling}). In principle, this voltage noise
will also induce a fluctuation of the gradient forces exerted on
the CNT [cf.~Eq.~(\ref{deflection})] resulting in additional
dissipation of the resonator mode. However this effect is
completely negligible for relevant parameters \footnote{A
  straightforward derivation using that the susceptibility
  associated to the voltage noise is determined by
  $\Im\{\chi(\omega)\}=R_J\omega$, yields for the corresponding
  contribution to the mechanical linewidth $\gamma_m\sim \frac{8
    R_J\left(V_2^2+V_1^2\right)}{\pi\sigma_GRL(V_2^2-V_1^2)^2}
  (\alpha_\parallel+\alpha_\perp)^2E_\perp(0)^2E_\parallel(z_0)^2$
  which for $R_J<1\mathsf{\Omega}$ implies
  $\gamma_m/2\pi\lesssim50\,\mu$Hz.}.

% (using that the imaginary part of the susceptibility associated to the
% voltage noise is given by $R_J\omega$)

Another potential source of exciton dephasing is the Brownian
motion of the suspended portion of the electrodes which entails
a limit to the suspended length, beyond which performance would
be degraded. The convenience of electrode suspension, beyond
that imposed by fabrication constraints, resides in the
suppression it can induce of the maximum electric field at the
neighbouring dielectric free surfaces ($E_\mathrm{d-f}$) with
respect to the maximum field. Assuming a cantilever geometry for
the suspended portion of the electrode and an
electrode-substrate gap size comparable to the suspended length
$L_e$, the suppression factor will scale as $(h_e/2L_e)^2$
---where $h_e$ is the transverse dimension of the electrode. For
the envisaged device (cf.~Fig.~\ref{fig:tips}) $h_e=30\,$nm and
$L_e=100\,$nm would already lead to
$E_\mathrm{d-f}\lesssim5\,$V$\mu^{-1}$. If we assume some
generic misalignment of the CNT, both the Brownian motion of the
electrode's flexural (in-plane and/or out-of-plane) and
compressional modes will contribute a term linear in the
relevant displacement $X_e$ to the corresponding fluctuation of
the effective confining potential $\delta
\hat{V}_\mathrm{eff}^{(B)}$. The latter will play a role
completely analogous to $\delta \hat{V}_\mathrm{eff}^{(V)}$ but
with $\delta \hat{V}_{1/2}$ replaced by the $\{X_e\}$. In all
cases the relevant fluctuation of the electric field satisfies
\begin{equation}\label{bound}
  \delta E_\parallel(z_0)\lesssim \frac{E_\parallel(z_0)}{z_0} X_e\,
\end{equation}
and a suitable model to estimate an upper bound for the
corresponding excess linewidth (FWHM) $\gamma_B$ is afforded by
considering only the contributions from the fundamental
resonances of each relevant branch (resonant frequency
$\omega_e$) which can be treated separately ---we assume in what
follows that the Brownian motions of different electrodes are
uncorrelated. These contributions can be analysed using the
exact solution of the ensuing independent-boson model
\cite{phdthesis}. If we assume Au electrodes with the
aforementioned cantilever dimensions, we have for the
fundamental flexural mode $\omega_e\sim1\,$GHz so that
$\Gamma\lesssim\omega_e$ (cf.~Secs.~\ref{subsec:laser-cool} and
\ref{subsec:strong-coupling}). There are then two distinct
effects of the electrodes' Brownian motion on the optical
response of the exciton that could potentially interfere with
our scheme: (i) generation of a sideband spectrum and (ii) pure
dephasing leading to an excess linewidth of the corresponding
ZPL. A critical parameter is the relative weight of the
single-phonon sidebands ($S$) which for each branch, in the
relevant regime $k_B T\gg\hbar\omega_e$ and using
Eqs.~(\ref{Veff}) and (\ref{bound}), can be bounded by
\begin{equation}\label{S}
  S_e\lesssim k_BT\left(\frac{16\pi\epsilon_0
      \epsilon_\parallel}{\hbar z_0}\right)^2
  \frac{\sigma_{eh}^6E_\parallel^4}{m_e\omega_e^4}
\end{equation}
where $m_e$ is the corresponding effective mass. If we consider
the envisaged device parameters
(cf.~Sec.~\ref{subsec:Xconfinement} and Fig.~\ref{fig:tips}),
the aforementioned ``electrode'' cantilever dimensions, the
material properties of Au and $T=4.2\,$K, we obtain from
Eq.~(\ref{S}) adding over all relevant branches of both
electrodes; $S\lesssim0.07$ ---the ratio of the corresponding
polaron shift to the one associated to the CNT resonator
($\eta^2\omega_0$) is $\lesssim0.03$
(cf.~Sec.~\ref{subsec:laser-cool}). Thus, the influence of (i)
on the proposed optomechanical manipulations would be minimal.

In turn, the pure dephasing (ii) is determined by the weak
dissipation of these resonances induced by the 3D substrate and
is characterised by an Ohmic environmental spectral density. The
conditions $S\ll1$ and $k_B T\gg\hbar\omega_e$ are sufficient to
ensure that the dephasing is Markovian and amenable to a
treatment similar to the one we have followed for the voltage
fluctuations. Each of the relevant fundamental resonances
coupled to the substrate satisfies \cite{Wilson-Rae08}
\begin{equation}\label{Xe}
\int_{-\infty}^{\infty}\langle X_e(t)X_e(0)\rangle\rmd t=
\left(\frac{2 x_e^{(0)}}{\omega_e}\right)^2\frac{k_BT}{\hbar Q_e}
\end{equation}
where $x_e^{(0)}$ and $Q_e$ are, respectively, the corresponding
zero-point motion and $Q$-value. Then, following a procedure
analogous to the one used for estimating $\gamma_p$, we obtain
from Eqs.~(\ref{Veff}), (\ref{bound}) and (\ref{Xe})
\begin{equation}\label{gammaB}%prefactor =1.2 for \nu_s=1/3
  \frac{\gamma_B}{2\pi}\lesssim k_BT
  \frac{\sqrt{\rho_s}}{E_s^{3/2}}\left(\frac{16\pi\epsilon_0
      \epsilon_\parallel}{\hbar z_0}\right)^2\sigma_{eh}^6E_\parallel^4\,.
\end{equation}
Here we have also assumed that given the low aspect ratio, the
$\{Q_e\}$ are limited by clamping losses, used the results of
Ref.~\cite{Wilson-Rae08} (Table I) which imply that
$m_vQ_v\omega_v^3=m_hQ_h\omega_h^3\sim m_cQ_c\omega_c^3\sim
E_s^{3/2}/\sqrt{\rho_s}$, and added the bounds for all three
relevant branches of both electrodes ---here $v$, $h$ and $c$
label, respectively, the flexural vertical, flexural horizontal
and compressional fundamental resonances of the ``electrode''
cantilever and the dimensionless prefactors are functions of
$\nu_s$. Strikingly once a cantilever geometry is assumed, the
dimensionless prefactor close to unity in Eq.~(\ref{gammaB}) is
independent of any other property of the electrode as it only
depends on the Poisson ratio of the substrate. Thus, the
magnitude of the suspended length $L_e$ would only begin to play
a role in exciton dephasing if it became large enough to ensure
that the mechanical dissipation were dominated by internal
losses. Finally, we find that for the proposed parameters
(cf.~Sec.~\ref{subsec:Xconfinement} and Fig.~\ref{fig:tips}) and
typical substrate properties determined by
$\rho_s=2\,$gcm$^{-3}$ and $E_s=100\,$GPa, Eq.~(\ref{gammaB})
yields a completely negligible excess linewidth
$\gamma_B/2\pi\lesssim20\,$kHz$\,\ll\Gamma$.

Lastly, we have estimated radiation pressure effects and found that
the corresponding static deflection and shot noise heating rate are
also completely negligible, with typical parameters implying a
launched power for a diffraction-limited spot size below $1\,$nW.

\subsection{Strong coupling}\label{subsec:strong-coupling}

Beyond optical transduction and cooling, the platform introduced
here could allow us to realise a mechanical analogue of the strong
and ultra-strong coupling regimes of cavity QED. If the driving
laser is tuned exactly on resonance, i.e. $\delta=0$ in
Eq.~(\ref{H_sys_original}), after a $\pi/2$ rotation of the
pseudospin around $\hat{y}$ ($\sigma_z\to\sigma_x$,
$\sigma_x\to-\sigma_z$), the system Hamiltonian reduces to the
% Jaynes-Cummings
Rabi model \cite{Braak11} ($H'_\mathrm{sys}\to H"_\mathrm{sys}$)
\begin{equation}
H"_\mathrm{sys}= -\frac{\Omega}{2}\sigma_z +
  \frac{\eta\omega_0}{2}\sigma_x\left(b_0^{\vphantom{\dag}}+b_0^{\dag}\right)+
  \omega_0 b_0^\dag b_0^{\vphantom{\dag}}%\,.
\end{equation}
with the spin degree of freedom afforded by the NTQD states
dressed by the laser field, i.e.
% \begin{equation}
%   \sigma_z|\pm\rangle=\pm|\pm\rangle\,.
% \end{equation}
\begin{equation}
  |\pm\rangle'=\frac{1}{\sqrt{2}}\left(|\uparrow\rangle'\pm
    |\downarrow\rangle'\right)\to|\downarrow/\uparrow\rangle"\,.
\end{equation}
Here $|\uparrow\rangle'\equiv|\psi_{00-}\rangle$ corresponds to the
relevant single-exciton state and $|\downarrow\rangle'$ to the empty
NTQD. The spin-oscillator coupling and resonance condition are given,
respectively, by $\eta\omega_0/2$ and $\Omega=\omega_0$. Thus, given
$1/Q\ll\eta\ll1$ reaching the strong coupling regime of the
Jaynes-Cummings model (i.e.~the Rabi model neglecting counter-rotating
terms) depends on satisfying $\eta\omega_0\gtrsim\Gamma$, which for
the aforementioned parameters would require $\Gamma/2\pi\lesssim
100\,$MHz. Here the precise threshold will differ from the more usual
one, $\eta\omega_0=\Gamma/2$, given that the dissipation of the
pseudospin induced by the radiation will no longer correspond to a
simple relaxation channel. As discussed in Sec.~\ref{sec:Xmodel} the
radiative contribution to $\Gamma$ can be tuned with the axial
magnetic field. We note that as the bright-dark exciton splitting lies
in the 100$\,$GHz range (cf.~Sec.~\ref{sec:Xmodel}) and the radiative
decay rate of the bright exciton is in the 10$\,$GHz range
\cite{Htoon04}, $\Gamma\sim 100\,$MHz with
$\Omega\sim\omega_0\sim1\,$GHz is compatible with keeping the
excitation of the bright exciton off-resonant.

The strong coupling regime discussed here is akin to the
ion-trap realisations of the Jaynes-Cummings model and the
parametric normal mode splitting in cavity optomechanics
\cite{Dobrindt08} in that the large energy-scale discrepancy
between the optical and mechanical domains is bridged by the
driving laser which entails an advantageous
frequency-upconversion of the output. It offers a wide range of
possibilities for the demonstration of quantum signatures in the
motion.  In particular a judicious modulation of $\eta$ locked
to pulsed laser excitation could allow us to emulate the adiabatic
passage scheme used in Ref.~\cite{Gleyzes07} for performing QND
measurements of the oscillator's energy.  This would enable the
observation of motional quantum jumps. 

Furthermore, from the analysis in Sec.~\ref{subsec:implementation} it
follows that higher transverse fields $\mathcal{E}_\perp$ could be
applied without disrupting the scheme. Then, a ten-fold increase in
$\mathcal{E}_\perp$ would result in $\eta\sim1$ allowing to crossover
to the ultra-strong coupling regime of the Rabi model which has not
been realised so far \cite{Braak11}. In this regime where the coupling
becomes comparable with or larger than the resonator's level spacing,
the ground state is characterised by an appreciable mean phonon
occupancy and presents substantial entanglement between the excitonic
quantum dot and the resonator \cite{Emary03}. To gain insight into
this regime it is instructive to consider the limit $\eta\gg1$ which
is simple to analyse in the ``polaronic'' representation by treating
the second term in the L.H.S. of Eq.~(\ref{spinboson_sys})
perturbatively. Thus, one obtains for the ground state of the coupled
system
\begin{equation}\label{cat}
  |\phi_0\rangle=\frac{1}{2}\left[|+\rangle'\left(|+\!\alpha\rangle
      +|-\!\alpha\rangle\right)+|-\rangle'\left(|+\!\alpha\rangle
      -|-\!\alpha\rangle\right)\right] 
\end{equation}
where
$|\pm\alpha\rangle=e^{\pm\frac{\eta\omega_0}{2}(b_0-b_0^\dag)}|0\rangle$.
 % Equation (\ref{cat})
This corresponds to a maximally entangled state for which
measurements of $\sigma_x$ yield orthogonal Schr\"{o}dinger
cat-states of the resonator.

\section{Conclusions and outlook}

In conclusion, we set forth a scheme for inducing strong
optomechanical effects in suspended CNTs via the deformation
potential exciton-phonon coupling. This provides an alternative
to radiation-pressure based schemes
\cite{Kippenberg08,Marquardt09,Favero09} for an ultra-low mass
and high frequency nanoscale resonator leading to large
backaction-cooling factors and opening a direct route to the
quantum behaviour of a ``macroscopic'' mechanical degree of
freedom as it affords a mechanical analogue of
cavity-QED % realization of the Jaynes-Cummings (and
% Rabi) model(s)
\cite{Armour02,Blencowe04}. Most importantly, these breakthroughs rely
on a lifetime-limited zero phonon line much narrower than the smallest
CNT linewidths reported so far \cite{Htoon04}. Indeed, the envisaged
NTQDs will allow us to suppress the two most likely
linewidth-broadening mechanisms, namely: inhomogeneous broadening and
phonon-induced dephasing \cite{Galland08I}, by providing a controlled
electrostatic environment and strong confinement of low-frequency
phonons. Beyond nanomechanical applications, a doped version of these
NTQDs will enable a tunable spin-photon interface
\cite{Galland08II}. Our technique could also apply to other high
quality one-dimensional semiconductors such as prismatic nanowires
\cite{Fontcuberta08}. Lastly, the simple approximation
[Eq.~(\ref{HamiltonianQD-ph})] we have derived for the deformation
potential of a confined $E_{11}$ direct exciton subjected to a
transverse electric field will further the understanding of
electron-phonon interactions in these CNT systems.

IWR acknowledges helpful discussions with Naser Qureshi, Adrian
Bachtold and Peter Rabl, and financial support from Nanosystems
Initiative Munich and DFG grant WI-3859/1-1.

\section*{References}
\bibliographystyle{iopart-num}%unsrt.bst

%\bibliography{anomalous}

\providecommand{\newblock}{}

\end{document}